\documentclass[english,prb,aps,showpacs,floatfix,amsmath,amsfonts,twocolumn]{revtex4-2}
\usepackage{lmodern}

\usepackage[T1]{fontenc}
\usepackage[latin9]{inputenc}
\usepackage{amsmath}
\usepackage{amssymb}
\usepackage{graphicx}
\usepackage{esint}
\usepackage{hyperref}

\makeatletter
\usepackage{epsfig}
\usepackage{dcolumn}
\usepackage[english]{babel}

\makeatother

\usepackage{babel}
\begin{document}
	\title{Plasmon polariton assisted second-harmonic generation in graphene}
	\author{Jo\~{a}o M. Alendouro Pinho$^{1}$, Sim\~{a}o S. Cardoso$^{2}$, Yuliy V. Bludov$^{2}$\email{bludov@fisica.uminho.pt}, Jo\~{a}o M. Viana Parente Lopes$^{1}$, Vladimir V. Konotop$^{3}$, Joel D. Cox$^{4,5}$, Nuno M.R. Peres$^{2,4,6}$}
	\affiliation{$^{1}$ Physics Center of Minho and Porto Universities (CF-UM-UP), Laboratory of Physics for Materials and Emergent Technologies (LaPMET), and Faculty of Science, University of Porto, Campo Alegre, 4169-007 Porto, Portugal}
	\affiliation{$^{2}$ Physics Center of Minho and Porto Universities (CF-UM-UP), Laboratory of Physics for Materials and Emergent Technologies (LaPMET), and Department of Physics, University of Minho, Campus of Gualtar, 4710-057 Braga, Portugal}
	\affiliation{$^{3}$Department of Physics, Center for Theoretical and Computational Physics, University of Lisbon, Campo Grande, C8, 1749-016 Lisbon, Portugal}
	\affiliation{$^{4}$POLIMA---Center for Polariton-driven Light--Matter Interactions, University of Southern Denmark, Campusvej 55, DK-5230 Odense M, Denmark}
	\affiliation{$^{5}$Danish Institute for Advanced Study, University of Southern Denmark, Campusvej 55, DK-5230 Odense M, Denmark}
	\affiliation{$^{6}$International Iberian Nanotechnology Laboratory (INL), Av. Mestre Jos\'{e} Veiga, 4715-330 Braga, Portugal}
	\begin{abstract}
		In this paper we present a theoretical examination of second-harmonic generation (SHG) in a graphene monolayer integrated within an attenuated total internal reflection (ATR) configuration. By embedding graphene in this optical setup, we explore the enhancement in the nonlinear optical response, particularly focusing on the efficiency of SHG.
		Our analysis reveals that the excitation of surface plasmon-polaritons (SPPs) plays a central role in significantly boosting the efficiency of SHG. The unique electronic properties of graphene, combined with the resonant characteristics of SPPs, create a synergistic effect that amplifies the nonlinear optical signals. This enhancement is attributed to the strong field confinement and the resonant nature of SPPs, which effectively increase the interaction between the incident light and the graphene monolayer.
		Furthermore, we analyze the underlying mechanisms that govern this process, providing a comprehensive theoretical framework that elucidates the interplay between graphene's electronic structure and the optical fields. Our findings suggest that the ATR scheme not only facilitates the excitation of SPPs but also optimizes the conditions for SHG.
	\end{abstract}
	\maketitle
	
	\section{Introduction}
	
	Beyond its many unusual and intriguing properties, graphene--a two-dimensional carbon monolayer--also demonstrates a relatively high nonlinear optical response \cite{2Dmat-review-Autere2018-advmat,2Dmat-review-Guo2019-lprev}. The large cubic nonlinearity in graphene, because of the conical dispersion of free electrons, was first predicted theoretically \cite{graphene-mikhailov2007-epl} and later confirmed experimentally in measurements of four-wave mixing \cite{graphene-exp-Hendry2010-prl} and third-harmonic generation \cite{graphene-exp-thg-Hong2013-prx,graphene-exp-thg-Kumar2013-prb}. At the same time, quadratic optical nonlinearity in graphene cannot be triggered by an external electromagnetic wave impinging normally on a graphene sheet owing to its centrosymmetric crystal structure. As a result, quadratic nonlinear optical phenomena in graphene necessitate mechanisms that break inversion symmetry \cite{gr-nonl-chi2-Cheng2015-prb}. Among a series of other possibilities (including curving or straining graphene \cite{gr-nonl-chi2-exp-Lin2014-apl,gr-nonl-chi2-exp-Lu2023-natcomm}; applied dc current \cite{gr-nonl-chi2-Cheng2014-oe,gr-bilay-nonl-chi2-Wu2012-nl,gr-nonl-chi2-Sekwao2015-apl,gr-nonl-chi2-Bykov2012-prb,gr-nonl-chi2-exp-An2013-nl}; valley polarization \cite{gr-nonl-chi2-Wehling2015-prb,gr-nonl-chi2-Golub2014-prb}; massive graphene, which is relevant for graphene on top of a substrate \cite{gr-nonl-chi2-Vandelli2019-prb}, and patterning graphene \cite{gr-nonl-plas-patt-Cox2014-ncomm,gr-nonl-plas-patt-Manzoni2015-njp,gr-nonl-plas-patt-Cox2017-prb}), the breaking of inversion symmetry can be achieved in the situation of excitation by optical fields with nonzero in-plane wavevector \cite{gr-nonl-chi2-Rapoport2017-jopt}, which can happen in the case of oblique incidence \cite{gr-nonl-chi2-exp-Dean2009-apl}. Nevertheless, in this situation, the nonlinear response is proportional to the in-plane momentum itself \cite{gr-nonl-chi2-Glazov2011-jetpl}, and thus is relatively weak in the general case.

	The nonlinear response of graphene can be enhanced considerably if graphene is coupled to some resonant system \cite{shg-rev-Huang2024-afm}, like an optical cavity \cite{gr-thg-cavity-Beckerleg2018-apl,gr-thg-cavity-Savostianova2015-apl,gr-thg-cavity-Zhao2019-optcomm,gr-nonl-shg-cavity-Song2023-oe} or photonic crystal \cite{gr-nonl-cavity-Wang2017-josab}. One possible way to effectively increase the magnitude of the in-plane wavevector component is to couple light with surface plasmon polaritons in graphene. Surface plasmon polaritons (SPPs), being surface waves, i.e., for which the electromagnetic field decays exponentially with increasing distance from graphene, possess higher in-plane wavevectors than bulk waves \cite{spp-gr-rev-Bludov2013-ijmfb}. Earlier it has been demonstrated that excitation of propagating plasmons \cite{gr-chi2-exp-plas-Lobet2015-nl,gr-nonl-hhg-plas-Calafell2022-aom,gr-thg-exp-Deinert2020-acsnano,gr-thg-exp-plas-Calafell2020-natnano} or localized ones \cite{gr-nonl-plas-patt-Cox2014-ncomm,gr-nonl-chi2-plas-loc-Smirnova2014-prb,gr-nonl-chi2-plas-loc-Aguillon2023-jpcl,gr-nonl-plas-loc-Cox2019-acccr} in graphene leads to considerable enhancement of various nonlinear optical phenomena. At the same time, in the case of quadratic nonlinearity, the inverse situation can take place \cite{gr-nonl-twm-exp-Constant2016-natphys}, where the excitation of surface plasmon polaritons in graphene from free space illumination was achieved by means of three-wave mixing.
	
	In this paper, we consider nonlinear light-matter interactions in a graphene monolayer incorporated into an attenuated total reflectance (ATR) structure. Specifically, we consider that a graphene sheet is deposited on a semi-infinite dielectric substrate and is excited by an evanescent wave, produced by total internal reflection at the boundary between vacuum and a dielectric semi-infinite prism. The paper is organized as follows. In Sec.\,\ref{sec:problem-statement}, we present particular solutions of Maxwell's equations in different spatial domains of layered media. In Sec.\,\ref{sec:2D-nonlinear-current}, we obtain an expression for the 2D nonlinear current using the Boltzmann transport equation formalism. In Sec.\,\ref{sec:Boundary-conditions}, we describe particular forms of boundary conditions, which, after being applied, give the nonlinear system of coupled equations describing second-harmonic generation (SHG) processes. In Sec.\,\ref{sec:Reflectance-and-SHG}, we demonstrate that the excitation of SPPs considerably enhances SHG efficiency. A brief conclusions section ends the central part of the paper. A set of appendices provide details of the calculations and an analysis of the best conditions for SHG in graphene.
	
	\section{Problem statement: fields in Layered Media}
	
	\label{sec:problem-statement}
\begin{figure}
		\includegraphics[width=8.5cm]{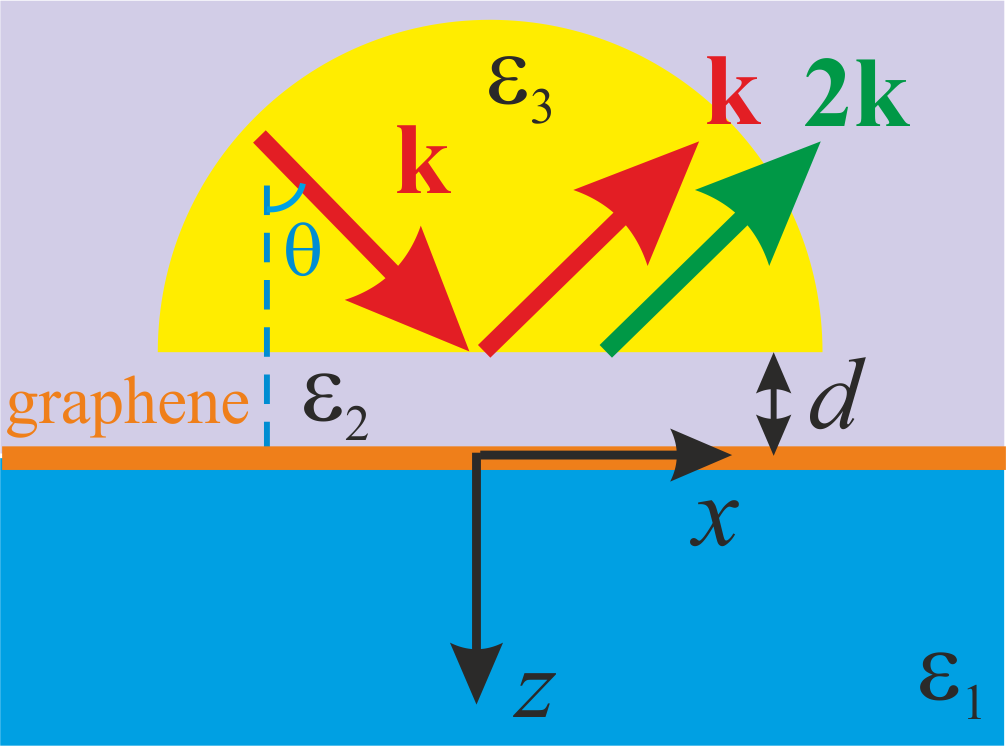}
		\caption{Schematics of the graphene-based ATR structure. Although dielectric prism is depicted semicircular, like in experiments, throughout the paper it is considered to have an infinite radius, i.e. occupies the whole half-space $z<-d$.}
		\label{fig:geometry}
	\end{figure}

We consider a flat interface between two semi-infinite dielectric media, one of which (with dielectric constant $\varepsilon_{2}$) occupies a finite space $-d<z<0$, while the other (with dielectric permeability $\varepsilon_{1}$) occupies the half-space $z>0$ (see schematics in Fig.~\ref{fig:geometry}). A 2D graphene sheet exhibiting a finite second-harmonic response is positioned at the interface $z=0$. In the ATR configuration, a prism with dielectric constant $\varepsilon_{3}$ is arranged in the half-space $z<-d$, and TM-polarized incident plane waves with angular frequency $\omega$ propagate in the positive $z$-direction and fall on the prism surface with an angle of incidence $\theta>\theta_{crit}$, where $\theta_{crit}=\arcsin\left(\sqrt{\max\left(\varepsilon_{1},\varepsilon_{2}\right)/\varepsilon_{3}}\right)$ is a critical angle for total internal reflection. In this case, the TM-polarized wave has electromagnetic field components $\boldsymbol{\mathcal{H}_m}=\left(0,\mathcal{H}_{m,y},0\right)$, $\boldsymbol{\mathcal{E}_m}=\left(\mathcal{E}_{m,x},0,\mathcal{E}_{m,z}\right)$, and $\boldsymbol{\mathcal{D}_m}=\left(\mathcal{D}_{m,x},0,\mathcal{D}_{m,z}\right)$, 
	where $m$ is the index of the medium occupied by the dielectric with relative dielectric permittivity $\varepsilon_{m}$. Since we are interested in the SHG process, we will seek the solutions of Maxwell's equations (\ref{eq:MaxEx})--(\ref{eq:MaxHy}), expanding the electromagnetic field up to the second-harmonic of the incident wave, i.e., $\boldsymbol{\mathcal{H}}_m\left(x,z,t\right)=\mathbf{H}_m^{(1)}\left(x,z\right)e^{-i\omega t}+\mathbf{H}_m^{(2)}\left(x,z\right)e^{-2i\omega t} + {\rm c.c.}$, $\boldsymbol{\mathcal{E}}_m\left(x,z,t\right)=\mathbf{E}_m^{(1)}\left(x,z\right)e^{-i\omega t}+\mathbf{E}_m^{(2)}\left(x,z\right)e^{-2i\omega t} + {\rm c.c.}$, $\boldsymbol{\mathcal{D}}_m\left(x,z,t\right)=\varepsilon_{0}\varepsilon\left(\omega\right)\mathbf{E}_m^{(1)}\left(x,z\right)e^{-i\omega t}+\varepsilon_{0}\varepsilon\left(2\omega\right)\mathbf{E}_m^{(2)}\left(x,z\right)e^{-2i\omega t} + {\rm c.c.}$, where $\mathbf{H}^{(n)}\left(x,z\right)$, $\mathbf{E}^{(n)}\left(x,z\right)$ are vector amplitudes of first $\left(n=1\right)$ and second $\left(n=2\right)$ harmonics of the electromagnetic field. The dielectric 
	permittivity $\varepsilon\left(\omega\right)$ generally depends on the frequency. However, to simplify the model,  focusing on the main phenomenon, below we consider  the relative dielectric permittivity as frequency-independent.
	
Owing to the linearity of the dielectric media $m=1,...,3$ Maxwell equations can be solved separately for the first and second harmonics. The particular solutions of Maxwell equations inside the semi-infinite prism ($m=3$) and substrate ($m=1$) can be represented in the form of field matrices, and parameterized via the amplitudes of incident and reflected waves. At the same time, solutions of Maxwell equation inside the finite dielectric medium $m=2$ can be represented in the form of a transfer matrix, and parameterized via the amplitudes of the tangential components of the electric and magnetic fields at the boundary \cite{trans-mat-Cheng2019-jpphot}. The particular forms of the solutions of Maxwell's equations are given in Appendix \ref{sec:Maxwell}.

	\section{Two-dimensional Nonlinear Current in Graphene}
	
	\label{sec:2D-nonlinear-current}
	
	In the previous section, all dielectric media are assumed to respond linearly to electric fields. As a result, Maxwell's equations (\ref{eq:MaxEx})--(\ref{eq:MaxHy}) in these media can be solved separately for first and second harmonics. At the same time, the nonlinear optical response of the graphene layer facilitates the interaction between first and second harmonics. In what follows, the current induced in the graphene layer will be characterized as a combination of linear, $\mathcal{J}_{x}^{(L)}\left(x,t\right)$, and nonlinear, $\mathcal{J}_{x}^{(NL)}\left(x,t\right)$, currents according to
	\begin{eqnarray}
		\mathcal{J}_{x}\left(x,t\right)=\mathcal{J}_{x}^{(L)}\left(x,t\right)+\mathcal{J}_{x}^{(NL)}\left(x,t\right).\label{eq:j-2D}
	\end{eqnarray}
	To obtain an expression for the current in graphene, we start from the Boltzmann equation for the non-equilibrium distribution function $f\left(\mathbf{r},t,\mathbf{q},\omega\right)$,
	\begin{align}
		\frac{\partial f\left(\mathbf{r},t,\mathbf{q},\omega\right)}{\partial t}+\frac{e}{\hbar}\boldsymbol{\mathcal{E}}_{\parallel}(\mathbf{r},t)\cdot\frac{\partial f\left(\mathbf{r},t,\mathbf{q},\omega\right)}{\partial\mathbf{q}}\label{eq:Boltzmann}
		\nonumber \\
		+\mathbf{v}(\mathbf{q})\cdot\frac{\partial f\left(\mathbf{r},t,\mathbf{q},\omega\right)}{\partial\mathbf{r}}+\gamma\left[f\left(\mathbf{r},t,\mathbf{q},\omega\right)-f_{0}(\mathbf{q})\right]=0,
		%,\nonumber
	\end{align}
	where $\mathbf{q}$ is the quasimomentum, $e$ is the elementary charge, $f_{0}\left(\mathbf{q}\right)$ is the Fermi-Dirac distribution, which for zero temperature will be written as $f_{0}\left(\mathbf{q}\right)=1-\Theta\left(q-q_{F}\right)$ [$\Theta\left(x\right)$ being the Heaviside step function, $q=\left|\mathbf{q}\right|=\sqrt{q_{x}^{2}+q_{y}^{2}}$, and $q_{F}$ is the Fermi wavevector], $\gamma$ is the inverse relaxation time, $\boldsymbol{\mathcal{E}}_{\parallel}(\mathbf{r},t)=\left(\mathcal{E}_{1,x}(x,z,t),0,0\right)$ is the electric field component parallel to graphene [the perpendicular component is not influenced owing to the 2D motion of charge carriers in graphene]. In further considerations, we admit the Dirac cone approximation for the energy spectrum of charge carriers in monolayer graphene, $\epsilon(\mathbf{q})=\hbar v_{F}q$ (with $v_{F}=10^{6}\,$m/s being the Fermi velocity). In this case, in Eq.~(\ref{eq:Boltzmann}), the group velocity of charge carriers $\mathbf{v}(\mathbf{q})=\hbar^{-1}\partial\epsilon(\mathbf{q})/\partial\mathbf{q}=v_{F}\mathbf{q}/q$. Also, we consider that graphene is doped by electrons, so charge $e<0$ and Fermi energy $E_{F}=\hbar v_{F}q_{F}>0$.
	
	Assuming that in 
	Eq.~(\ref{eq:Boltzmann}), the electric field is considered as a small perturbation, the solution can be represented as an expansion into a series $f\left(\mathbf{r},t,\mathbf{q},\omega\right)=f_{0}\left(\mathbf{r},t,\mathbf{q},\omega\right)+f_{1}\left(\mathbf{r},t,\mathbf{q},\omega\right)+f_{2}\left(\mathbf{r},t,\mathbf{q},\omega\right)+...$ (where $f_{2}\left(\mathbf{r},t,\mathbf{q},\omega\right)\ll f_{1}\left(\mathbf{r},t,\mathbf{q},\omega\right)\ll f_{0}\left(\mathbf{q}\right)$, and $f_{n}\left(\mathbf{r},t,\mathbf{q},\omega\right)\propto\left|\mathbf{\mathcal{E}}\right|^{n}$, as shown in Appendix \ref{sec:Ab-initio-calculation}). Equating the terms of the same order in the electric field leads to the set of equations
	\begin{eqnarray*}
		\frac{\partial f_{n}\left(\mathbf{r},t,\mathbf{q},\omega\right)}{\partial t}-\frac{|e|}{\hbar}\boldsymbol{\mathcal{E}}_{\parallel}(\mathbf{r},t)\cdot\frac{\partial f_{n-1}\left(\mathbf{r},t,\mathbf{q},\omega\right)}{\partial\mathbf{q}}\\
		+\mathbf{v}\cdot\frac{\partial f_{n}\left(\mathbf{r},t,\mathbf{q},\omega\right)}{\partial\mathbf{r}}+\gamma\left[f_{n}\left(\mathbf{r},t,\mathbf{q},\omega\right)-f_{0}(\mathbf{q})\right]=0,
	\end{eqnarray*}
	the solution of which can be represented as
	\begin{eqnarray}
		\label{eq:ite_solution_f}
		f_{n}(x,t,\mathbf{q},\omega)=\frac{|e|}{\hbar}\int_{-\infty}^{t}dt^{\prime}
		\exp[-\gamma\left(t-t^{\prime}\right)] 
		\nonumber \\
		\times\mathcal{E}_{1,x}\left(x-v_{x}\left(t-t^{\prime}\right),0,t^{\prime}\right)
		\nonumber \\
		\times\frac{\partial f_{n-1}\left(x-v_{x}\left(t-t^{\prime}\right),t^{\prime},\mathbf{q},\omega\right)}{\partial q_{x}}, 
	\end{eqnarray}
	for an electric field polarized along the $x$ axis. Equation (\ref{eq:ite_solution_f}) can be used as a bottom-up approach to obtain $f_{n}(x,t)$ iteratively. Therefore, using the fact that
	\begin{eqnarray}
		\frac{\partial f_{0}\left(\mathbf{q}\right)}{\partial q_{x}}=-\delta\left(q-q_{F}\right)\frac{q_{x}}{q},\label{eq:df0-dqx}
	\end{eqnarray}
	and the expression for the electric field (\ref{eq:EH1-mat}), the first-order correction $f_{1}(x,t)$ yields
	\begin{align}
		\label{eq:f1_exact_sol}
		f_{1}\left(x,t,\mathbf{q},\omega\right)=-\frac{|e|}{\hbar}
		\sum_{n=1}^{2}E_{1,x}^{\left(n\right)}\frac{\exp\left[in(k_{x}x-\omega t)\right]}{in(k_{x}v_{x}-\omega)+\gamma} 
		\nonumber \\
		\times\frac{q_{x}}{q}\delta\left(q-q_{F}\right)+{\rm c.c.} 
	\end{align}
	For small $k_{x}v_{F}\ll\omega$ (beyond the region of Landau damping), 
	Eq.~(\ref{eq:f1_exact_sol}) can be expanded into a Taylor series as
	\begin{align}
		\label{eq:f1-final}
		f_{1}\left(x,t,\mathbf{q},\omega\right)=-\frac{|e|}{\hbar}\sum_{n=1}^{2}E_{1,x}^{\left(n\right)}\frac{\exp\left[in(k_{x}x-\omega t)\right]}{\gamma-in\omega} 
		 \nonumber \\
		\times\left(1-\frac{ink_{x}v_{x}}{\gamma-in\omega}\right)\frac{q_{x}}{q}\delta\left(q-q_{F}\right)+{\rm c.c.}  
	\end{align}
	
	The first-order correction defines the linear current as
	\begin{eqnarray}
		\mathcal{J}_{x}^{(L)}\left(x,t\right)=-\frac{4\left|e\right|}{\left(2\pi\right)^{2}}\int d\mathbf{q}v_{x}f_{1}\left(x,t,\mathbf{q},\omega\right),\label{eq:jl}
	\end{eqnarray}
	where the factor 4 comes from spin and valley degeneration. In this case, substituting Eq.~(\ref{eq:f1-final}) into Eq.~(\ref{eq:jl}) and integrating, we obtain the linear contribution in the electric field due to linear conductivity
	\begin{align}
		\mathcal{J}_{x}^{(L)}\left(x,t\right)=\sum_{n=1}^{2}\sigma_{xx}\left(n\omega\right)E_{x}^{(n)}\label{eq:j-l}
		%\exp\left[ink_{x}x-in\omega t\right]
		e^{in(k_{x}x-\omega t)} 
		{\rm c.c.}  
	\end{align}
	where
	\begin{eqnarray*}
		\sigma_{xx}\left(\omega\right)=\frac{e^{2}E_{F}}{\pi\hbar^{2}}\frac{1}{\gamma-i\omega}
	\end{eqnarray*}
	is the Drude linear conductivity of graphene. The nonlinear current is defined by the correction of the second order $f_{2}\left(x,t,\mathbf{q},\omega\right)$, which can be obtained by differentiating the first-order correction Eq.~(\ref{eq:f1-final}),
	\begin{align}
		\label{eq:df1-dqx}
		%\begin{aligned}
			\frac{\partial f_{1}\left(x,t,\mathbf{q},\omega\right)}{\partial q_{x}} & =-\frac{|e|}{\hbar}\sum_{n=1}^{2}E_{1,x}^{\left(n\right)}\Gamma_{n}\left(\mathbf{q}\right)
		%\end{aligned} 
		%\\ \times \exp\left[ink_{x}x-in\omega t\right]
		e^{in(k_{x}x-\omega t)}+c.c., 
	\end{align}
	where 
	\begin{eqnarray*}
	\Gamma_{n}\left(\mathbf{q}\right)=\frac{\Upsilon\left(\mathbf{q}\right)}{\gamma-in\omega}-ink_{x}v_{F}\frac{\Phi\left(\mathbf{q}\right)}{\left(\gamma-in\omega\right)^{2}},\nonumber\\
		\Upsilon\left(\mathbf{q}\right)=\frac{q_{y}^{2}}{q^{3}}\delta\left(q-q_{F}\right)+\frac{q_{x}^{2}}{q^{2}}\frac{\partial\delta\left(q-q_{F}\right)}{\partial q},\\
\Phi\left(\mathbf{q}\right)=2\frac{q_{x}q_{y}^{2}}{q^{4}}\delta\left(q-q_{F}\right)+\frac{q_{x}^{3}}{q^{3}}\frac{\partial\delta\left(q-q_{F}\right)}{\partial q}.
	\end{eqnarray*}
	Substituting this result into 
	%Equation 
	Eq.~(\ref{eq:ite_solution_f})  we obtain
	\begin{equation}
		f_{2}(x,t,\mathbf{q},\omega)=\sum_{n=1}^{2}f_{2}^{(n)}(x,t,\mathbf{q},\omega)+{\rm c.c.} \label{eq:f2}
	\end{equation}
	Here
\begin{align*}
%	f_{2}^{(n)}(x,t,\mathbf{q},\omega)=-\frac{e^{2}}{\hbar^{2}}K^{(n)}\frac{\exp\left[in(k_{x}x-\omega t)\right]}{\gamma-in\omega}\\
%	\times\left(1-\frac{ink_{x}v_{x}}{\gamma-in\omega}\right),
%\\
f_{2}^{(n)}(x,t,\mathbf{q},\omega)=-\frac{e^{2}}{\hbar^{2}} \frac{K^{(n)} }{\gamma-in\omega}  \left(1-\frac{ink_{x}v_{x}}{\gamma-in\omega}\right)e^{in(k_{x}x-\omega t)},		
		\\
		K^{(1)}=E_{1,x}^{(2)}\overline{E_{1,x}^{(1)}}\left(\overline{\Gamma_{1}\left(\mathbf{q}\right)}+\Gamma_{2}\left(\mathbf{q}\right)\right),\\
		K^{(2)}=\left(E_{1,x}^{(1)}\right)^{2}\Gamma_{1}\left(\mathbf{q}\right),
\end{align*}
	and hereafter an overbar stands for complex conjugation.
	The nonlinear current can be calculated by integrating the second-order correction in the momentum plane (details can be found in Appendix \ref{sec:Ab-initio-calculation}),
	\begin{eqnarray}
		\mathcal{J}_{x}^{(NL)}\left(x,t\right)=-\frac{4\left|e\right|}{\left(2\pi\right)^{2}}\int d\mathbf{q}v_{x}f_{2}\left(x,t,\mathbf{q},\omega\right),\label{eq:j-nl}
	\end{eqnarray}
	and in explicit form expressed as
	\begin{align}
		\label{eq:j-nl-final}
		\mathcal{J}_{x}^{(NL)}\left(x,t\right)=\sigma_{xxx}^{(21)}E_{1,x}^{(2)}\overline{E_{1,x}^{(1)}}\exp\left[ik_{x}x-i\omega t\right]
		\nonumber \\
		+\sigma_{xxx}^{(11)}\left(E_{1,x}^{(1)}\right)^{2}\exp\left[2ik_{x}x-2i\omega t\right]+{\rm c.c.}, 
	\end{align}
	where
	\begin{align}
		\label{eq:sigma-11}
	 	\sigma_{xxx}^{(11)}=ik_{x}\frac{\left|e\right|^{3}v_{F}^{2}}{4\pi\hbar^{2}}\frac{1}{\gamma-i\omega}\frac{1}{\gamma-2i\omega}\left(\frac{1}{\gamma-i\omega}+\frac{4}{\gamma-i2\omega} \right)
	\end{align}
	and
	\begin{align}
		 \label{eq:sigma-21}
		 \sigma_{xxx}^{(21)}=&ik_{x}\frac{\left|e\right|^{3}v_{F}^{2}}{4\pi\hbar^{2}}
		 \frac{1}{\gamma-i\omega}\left[-\frac{1}{\left(\gamma+i\omega\right)^{2}}+\frac{2}{\gamma-i\omega}\frac{1}{\gamma+i\omega}
		 \right. \nonumber \\ 
		 &\left.
		 +\frac{2}{\left(\gamma-i2\omega\right)^{2}}+\frac{2}{\gamma-i\omega}\frac{1}{\gamma-i2\omega}\right],
	\end{align}
	%\begin{widetext}
	%where	
	%	\begin{eqnarray}
	%		\label{eq:sigma-21}
	%		\sigma_{xxx}^{(21)}=ik_{x}\frac{\left|e\right|^{3}v_{F}^{2}}{4\pi\hbar^{2}}\frac{1}{\gamma-i\omega}\left[-\frac{1}{\left(\gamma+i\omega\right)^{2}}+\frac{2}{\gamma-i\omega}\frac{1}{\gamma+i\omega}+\frac{2}{\left(\gamma-i2\omega\right)^{2}}+\frac{2}{\gamma-i\omega}\frac{1}{\gamma-i2\omega}\right],
	%		\\
	%		\sigma_{xxx}^{(11)}=ik_{x}\frac{\left|e\right|^{3}v_{F}^{2}}{4\pi\hbar^{2}}\frac{1}{\gamma-2i\omega}\left[\frac{1}{\left(\gamma-i\omega\right)^{2}}+\frac{4}{\gamma-i2\omega}\frac{1}{\gamma-i\omega}\right].\label{eq:sigma-11}
	%	\end{eqnarray}
	%\end{widetext}
	are nonlinear conductivities.
	 Notice that both nonlinear conductivities in Eqs.~(\ref{eq:sigma-11}) and (\ref{eq:sigma-21}) are proportional to the in-plane component of the wavevector $k_{x}$. In other words, these nonlinear conductivities are nonzero only for the case of oblique incidence, when the inversion symmetry is broken. The expression for the conductivity (\ref{eq:sigma-11}) corresponds to second-harmonic generation, and is in agreement with the results reported in Refs.~\cite{gr-nonl-plas-patt-Cox2017-prb,chi2-Cheng2017-scirep,chi2-gr-Wang2016-prb,rasmussen2023nonlocal} for a purely intraband (free electron) response at energies $\hbar\omega\lesssim E_{\rm F}$. Equation (\ref{eq:sigma-21}) describes a cascaded nonlinear optical process~\cite{rasmussen2023nonlocal}, where by the generated second-harmonic field oscillating as $e^{-2i\omega t}$ mixes with the counterrotating component of the fundamental field oscillating with $e^{i\omega t}$.
	
	\section{Boundary Conditions and Nonlinear System of Equations}
	
	\label{sec:Boundary-conditions}Boundary conditions across the interface can be expressed in the following form. At the interface $z=-d$ (without graphene), both tangential components of electromagnetic fields are continuous across the interface,
	\begin{eqnarray}
		\left(\begin{array}{c}
			{\cal E}_{2,x}(x,-d,t)\\
			Z_{0}{\cal H}_{2,y}(x,-d,t)
		\end{array}\right)=\left(\begin{array}{c}
			{\cal E}_{3,x}(x,-d,t)\\
			Z_{0}{\cal H}_{3,y}(x,-d,t)
		\end{array}\right).\label{eq:bc-md}
	\end{eqnarray}
	These boundary conditions allow one to obtain the electromagnetic field components at $z=0$, $\mathcal{E}_{2,x}\left(x,0,t\right)$ and $\mathcal{H}_{2,y}\left(x,0,t\right)$, from the consecutive applications of Eqs. (\ref{eq:EH3-mat}), (\ref{eq:bc-md}), and (\ref{eq:EH2-mat}),
	\begin{align}
		\label{eq:EH20-mat}
		\left(\begin{array}{c}
			{\cal E}_{2,x}(x,0,t)\\
			Z_{0}{\cal H}_{2,y}(x,0,t)
		\end{array}\right)=\sum_{n=1}^{2}\hat{F}_{tot}^{(n)}\left(\begin{array}{c}
			\delta_{n,1}E_{i}\\
			E_{3,x}^{(n)}
		\end{array}\right) 
		\nonumber 
		\\
		\times\exp\left[ink_{x}x-in\omega t\right]+{\rm c.c.}, 
	\end{align}
	with
	\begin{eqnarray*}
		\hat{F}_{tot}^{(n)}=\hat{Q}_{2}^{(n)}\left(0\right)\hat{F}_{3}^{(n)}
	\end{eqnarray*}
	being the total field matrix of the $n$th harmonics. At the interface $z=0$, where graphene is present, the tangential component of the electric field is also continuous across the interface,
	\begin{eqnarray}
		\mathcal{E}_{2,x}\left(x,0,t\right)=\mathcal{E}_{1,x}\left(x,0,t\right),\label{eq:bc-e0}
	\end{eqnarray}
	but the tangential component of the magnetic field is discontinuous across the interface owing to the presence of the two-dimensional current in graphene,
	\begin{eqnarray}
		\mathcal{H}_{1,y}\left(x,0,t\right)-\mathcal{H}_{2,y}\left(x,0,t\right)=-\mathcal{J}_{x}\left(x,t\right),\label{eq:bc-h0}
	\end{eqnarray}
	where the 2D current is given by Eq.~(\ref{eq:j-2D}). Substitution of the upper lines of Eqs.~(\ref{eq:EH20-mat}) and (\ref{eq:EH1-mat}) into boundary conditions (\ref{eq:bc-e0}) results in the following relation between the amplitudes:
	\begin{align}
		\label{eq:bc-Ex}
		\left(\hat{F}_{tot}^{(n)}\right)_{11}E_{i}\delta_{n,1}+
		\left(\hat{F}_{tot}^{(n)}\right)_{12}E_{3,x}^{(n)} 
		%\\
		=\left(\hat{F}_{1}^{(n)}\right)_{11}E_{1,x}^{(n)}.
	\end{align}
	At the same time, substituting the lower lines in Eqs.~(\ref{eq:EH20-mat}) and (\ref{eq:EH1-mat}) into boundary condition (\ref{eq:bc-h0}), and taking into account Eq.~(\ref{eq:bc-Ex}), we obtain a system of equations that governs the first and second harmonics of the electric field's in-plane components $E_{1,x}^{(1)}$ and $E_{1,x}^{(2)}$:
	\begin{eqnarray}
		\label{eq:E1x-final}
		D^{(1)}(d)E_{1,x}^{(1)}+Z_{0}\sigma_{xxx}^{(21)}E_{1,x}^{(2)}\overline{E_{1,x}^{(1)}}=GE_{i}, 
		\\
		\label{eq:E2x-final}
		D^{(2)}(d)E_{1,x}^{(2)}+Z_{0}\sigma_{xxx}^{(11)}\left(E_{1,x}^{(1)}\right)^{2}=0, 
	\end{eqnarray}
	where
	\begin{align}
		\label{eq:Dn}
		D^{(n)}(d)&=\left(\hat{F}_{1}^{(n)}\right)_{21}-\left(\hat{F}_{tot}^{(n)}\right)_{22}\frac{\left(\hat{F}_{1}^{(n)}\right)_{11}}{\left(\hat{F}_{tot}^{(n)}\right)_{12}} 
		%\nonumber \\
		+Z_{0}\sigma_{xx}\left(n\omega\right),
		 \\
		G&=\left(\hat{F}_{tot}^{(1)}\right)_{21}-
		\left(\hat{F}_{tot}^{(1)}\right)_{22}\frac{\left(\hat{F}_{tot}^{(1)}\right)_{11}}{\left(\hat{F}_{tot}^{(1)}\right)_{12}}. 
	\end{align}
	
	\section{Reflectance and SHG Efficiency}
	\label{sec:Reflectance-and-SHG}
	
	In order to obtain the reflectance and transmittance coefficients, we start from the Poynting vector. In general form, the Poynting vector, averaged in time, will be written as $\boldsymbol{S}=\mathrm{Re}\left\{ \boldsymbol{E}\times\overline{\boldsymbol{H}}\right\} /2$. Here, 
	%the overbar stands for complex conjugation, and 
	$\times$ stands for the vector product. Since the direction normal to interfaces is the $z$ axis, the reflectance and transmittance coefficients will be proportional to the $z$ components of the Poynting vector, i.e.,
	\begin{eqnarray}
		S_{z}=E_{x}\overline{H_{y}}/2.\label{eq:Sz}
	\end{eqnarray}
	For the incident wave, the respective components of the electromagnetic field can be obtained from Eq.~(\ref{eq:EH3-mat-alt}), and the $z$ component of the Poynting vector will be $S_{z,i}=E_{i}\overline{H_{i}}/2=\left|E_{i}\right|^{2}\sqrt{\varepsilon_{3}}/\left(2Z_{0}\cos\theta\right)$. The Poynting vector of both harmonics of the reflected wave can be obtained from the respective in-plane components of the electromagnetic field in Eq.~(\ref{eq:EH3-mat-alt}) and represented as $S_{3,z}^{(n)}=E_{3,x}^{(n)}\overline{H_{3,y}^{(n)}}/2=-\left|E_{3,x}^{(n)}\right|^{2}\sqrt{\varepsilon_{3}}/\left(2Z_{0}\cos\theta\right)$. Here, the negative sign means the energy flux in the negative direction of the $z$ axis, which corresponds to the backward-propagating wave. For the transmitted wave (in the semi-infinite medium with permittivity $\varepsilon_{1}$), the situation turns out to be more complicated. The respective tangential components of the electromagnetic field [see Eq.\,(\ref{eq:EH1-mat-alt})], being substituted into the general expression for the $z$ component of the Poynting vector (\ref{eq:Sz}), give
	\begin{eqnarray*}
		&&S_{1,z}^{(n)}=E_{1,x}^{(n)}\overline{H_{1,y}^{(n)}}/2\\
		&&=\left|E_{1,x}^{(n)}\right|^{2}\mathrm{Re}\left\{ \varepsilon_{1}/\left(2Z_{0}\sqrt{\varepsilon_{1}-\varepsilon_{3}\sin^{2}\theta}\right)\right\}.
	\end{eqnarray*}
	In the above-mentioned case of the ATR structure, since the angle of incidence $\theta$ is bigger than the critical angle, $\sqrt{\varepsilon_{1}-\varepsilon_{3}\sin^{2}\theta}$ is a purely imaginary value, and, hence, $S_{1,z}^{(n)}\equiv0$ (this property stems from the fact that the evanescent wave does not transfer energy). Nevertheless, for the case $\varepsilon_{3}<\varepsilon_{1}$ (this case will also be considered further in the paper), the wave in the substrate turns out to be propagating, and $S_{1,z}^{(n)}\ne0$.
	Thus, the reflectance $R^{(1)}$ and transmittance $T^{(1)}$ of the first harmonics will be defined as
    \begin{eqnarray*}
    	R^{(1)}=-\frac{S_{3,z}^{(1)}}{S_{z,i}},\qquad T^{(1)}=\frac{S_{1,z}^{(1)}}{S_{z,i}},
    \end{eqnarray*}
	while the respective SHG efficiencies are
	\begin{eqnarray*}
			R^{(2)}=-\frac{S_{3,z}^{(2)}}{S_{z,i}},\qquad T^{(2)}=\frac{S_{1,z}^{(2)}}{S_{z,i}}.%
	\end{eqnarray*}
	\begin{figure}
		\includegraphics[width=8.5cm]{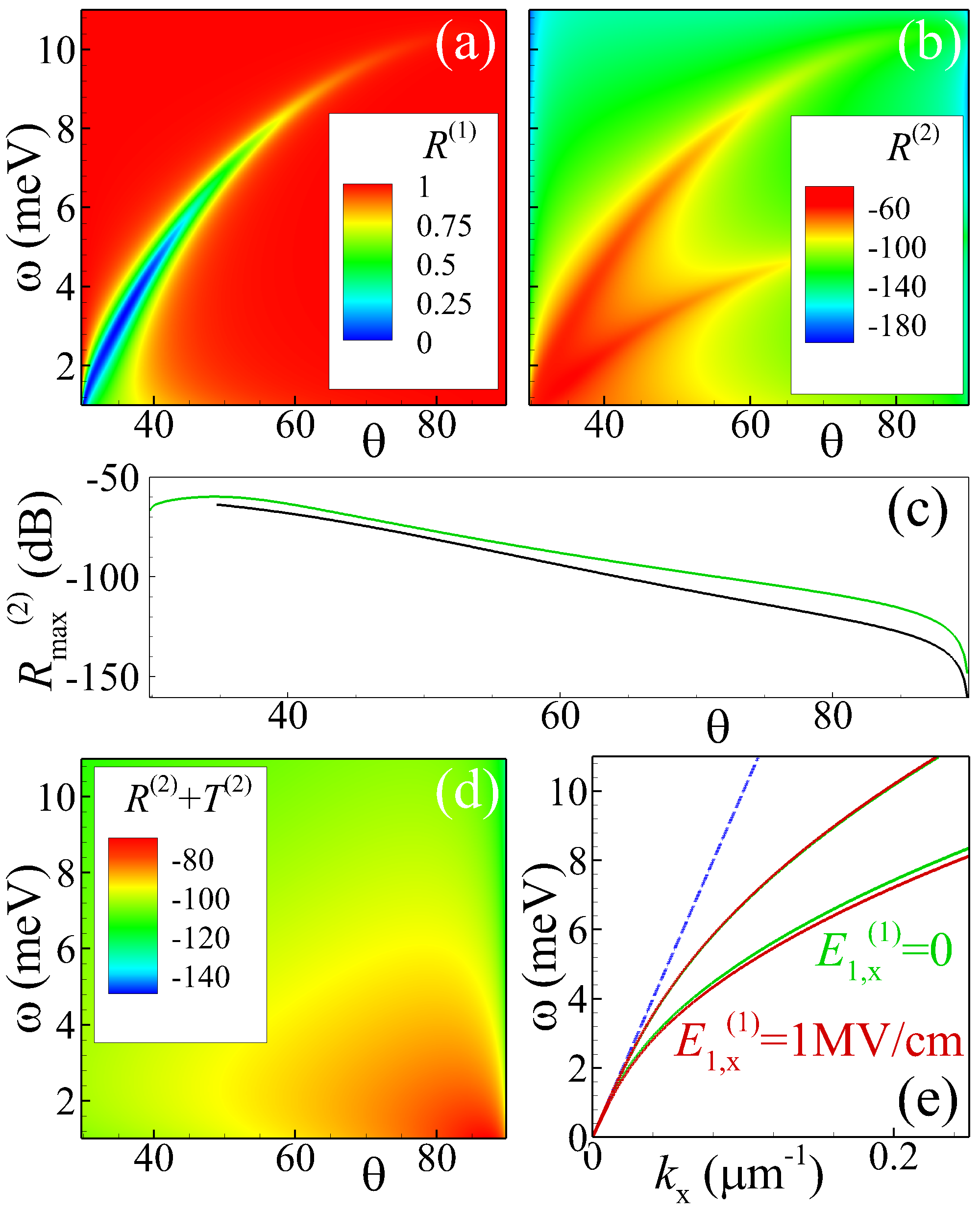}
		\caption{(a)(b) Reflectance $R^{(1)}$ (a) and SHG efficiency $R^{(2)}$ (b) versus angle of incidence $\theta$ and frequency $\omega$ of ATR structure with parameters: $d=10\,\mu$m, $\varepsilon_{3}=16$, $\varepsilon_{2}=1$, $\varepsilon_{1}=3.9$, $E_{F}=0.5\,$eV, $\gamma=0.25\,$meV, $E_{i}=1\,$kV/cm. (c) Maximal values of SHG efficiency $R^{(2)}_{max}$ vs angle of incidence $\theta$ for high-frequency (green line) and low-frequency modes (black lines) for the same parameters as those in (b). (d) SHG efficiency $R^{(2)}+T^{(2)}$ vs angle of incidence $\theta$ and frequency $\omega$ for the same structure as in (b), but with $\varepsilon_{3}=1$. (e) Frequency vs in-plane component of wavevector $k_{x}$ of nonlinear eigenmodes [Eq.~\ref{eq:dr} with parameters $\varepsilon_{2}=1$, $\varepsilon_{1}=3.9$, $\gamma=0$, $E_{F}=0.5\,$eV, $E_{1,x}^{(1)}=0$ (green lines), or $E_{1,x}^{(1)}=1\,$MV/cm (red lines)].}
		\label{fig:reflectance}
	\end{figure}
	The reflectance of the ATR structure as a function of frequency and angle of incidence is shown in Fig.~\ref{fig:reflectance}(a). As it can be seen, the reflectance of the ATR structure with the nonlinear graphene has minima at certain frequencies and angles of incidence, owing to the excitation of SPPs in graphene. This phenomenon is similar to that observed in linear graphene \cite{atr-Bludov2010-epl}. The reflectance minimum occurs when the frequency and in-plane wavevector of the incident wave match those of the SPP eigenmode: In this case, the energy of the incident wave is effectively transferred into that of the SPP, thus reducing the intensity of the reflected wave. In the case of nonlinear graphene, the nonzero in-plane wavevector $k_{x}\ne0$ imposes quadratic nonlinearity in graphene owing to the breaking of inverse symmetry, which in its turn leads to the possibility of SHG. The respective SHG efficiency, depicted in Fig.~\ref{fig:reflectance}(b), is characterized by the presence of two maxima, whose physical origins will be explained further in the paper. The position of one of the maxima (high-frequency) almost coincides with the reflectance minima [compare Figures \ref{fig:reflectance}(a) and \ref{fig:reflectance}(b)], thus the excitation of SPPs in the system leads to a considerable increase of SHG efficiency [up to $\sim-59\,$dB for the parameters of Figs.~\ref{fig:reflectance}(b) and \ref{fig:reflectance}(c) for the low-frequency peak (green line)]. In particular, the SHG efficiency of the low-frequency peak is higher than that of the high-frequency peak [compare green and black lines in Fig.\,\ref{fig:reflectance}(c)]. For comparison, Fig.~\ref{fig:reflectance}(d) presents the SHG efficiency for the scheme without a prism ($\varepsilon_{3}=1$), where the incident propagating wave falls directly on the graphene. In this case, the transmitted wave in the substrate is also propagating, thus here the SHG efficiency is defined as $R^{(2)}+T^{(2)}$. In this type of structure, SPPs are not excited, and, as a result, the SHG efficiency is considerably less than in the ATR structure [Fig.~\ref{fig:reflectance}(b)].
	
	In order to explain the origins of the two maxima in Fig.~\ref{fig:reflectance}(b), we consider SPP eigenmodes sustained by nonlinear graphene. Since in the ATR structure electromagnetic waves in media $m=1,2$ are evanescent in the $z$ direction, $k_{m,z}$ are purely imaginary and can be parameterized as $k_{m,z}=ip_{m}$ with decay factor $p_{m}=\sqrt{k_{x}^{2}-\omega^{2}\varepsilon_{m}/c^{2}}$ being purely real, $\mathrm{Im}(p_{m})\equiv0$. In the limit of infinite thickness $\left(d\to\infty\right)$ of medium $m=2$, the parameters defined by Eq.~(\ref{eq:Dn}) acquire the form
	\begin{eqnarray}
		&&D^{(n)}\left(\infty\right)=\lim_{d \to \infty} D^{(n)}\left(d\right)\nonumber\\
		&&=\frac{\varepsilon_{1}\omega}{icp_{1}}+\frac{\omega\varepsilon_{2}}{icp_{2}}+Z_{0}\sigma_{xx}\left(n\omega\right).
	\end{eqnarray}
	Under this condition, and when the incident wave is absent, $E_{i}\equiv0$, Eqs.~(\ref{eq:E1x-final}) and (\ref{eq:E2x-final}) are represented as
	\begin{eqnarray}
		D^{(1)}\left(\infty\right)E_{1,x}^{(1)}+Z_{0}\sigma_{xxx}^{(21)}E_{1,x}^{(2)}\overline{E_{1,x}^{(1)}}=0,\label{eq:E1x-eigen}\\
		D^{(2)}\left(\infty\right)E_{1,x}^{(2)}+Z_{0}\sigma_{xxx}^{(11)}\left(E_{1,x}^{(1)}\right)^{2}=0.\label{eq:E2x-eigen}
	\end{eqnarray}
	This system of equations describes two-wave mixing \cite{2wm-Stegeman1996-oqe,2wm-Trillo1992-ol}, and determines the dispersion relation for SPP eigenmodes,
	\begin{eqnarray}
		D^{(1)}\left(\infty\right)D^{(2)}\left(\infty\right)-Z_{0}^{2}\sigma_{xxx}^{(21)}\sigma_{xxx}^{(11)}\left|E_{1,x}^{(1)}\right|^{2}=0.\label{eq:dr}
	\end{eqnarray}
	This nonlinear dispersion relation is parameterized by the in-plane component of the first-harmonic electric field $E_{1,x}^{(1)}$. In the linear limit $E_{1,x}^{(1)}=0$, the dispersion relation (\ref{eq:dr}) is split into two ones,
	\begin{eqnarray}
		\frac{\varepsilon_{1}\omega}{cp_{1}}+\frac{\omega\varepsilon_{2}}{cp_{2}}+iZ_{0}\sigma_{xx}\left(\omega\right)=0,\label{eq:dr1}\\
		\frac{\varepsilon_{1}\omega}{cp_{1}}+\frac{\omega\varepsilon_{2}}{cp_{2}}+iZ_{0}\sigma_{xx}\left(2\omega\right)=0,\label{eq:dr2}
	\end{eqnarray}
	Here, the first line \eqref{eq:dr1} represents the dispersion relation for linear SPPs in the graphene layer at frequency $\omega$ (see, e.g., Ref. \cite{spp-gr-rev-Bludov2013-ijmfb} for details), while the second line \eqref{eq:dr2} is the dispersion relation for the linear SPPs in graphene at doubled frequency $2\omega$ (note that a similar phenomenon---the presence of two resonances---was reported in Ref.~\cite{nonl-plas-Rostami2017-prb}). In the linear case, these two eigenmodes are decoupled, and their dispersion curves are depicted in Fig.~\ref{fig:reflectance}(d) by green solid lines where high- and low-frequency modes are determined by Eqs.~(\ref{eq:dr1}) and (\ref{eq:dr2}), respectively [below they are referred to as single- and double-frequency surface plasmon-polariton (SFSPP and DFSPP)]. For nonzero in-plane electric field $E_{1,x}^{(1)}\ne0$, these two SPP modes are coupled, and the dispersion relation (\ref{eq:dr}) turns out to be nonlinear. The respective eigenmodes are depicted in Fig.~\ref{fig:reflectance}(d) by red solid lines. From a comparison of Fig.~\ref{fig:reflectance}(b) and Fig.~\ref{fig:reflectance}(d), one can conclude that two maxima of SHG efficiency are determined by resonant excitation of coupled eigenmodes.
	
	\begin{figure}
		\includegraphics[width=8.5cm]{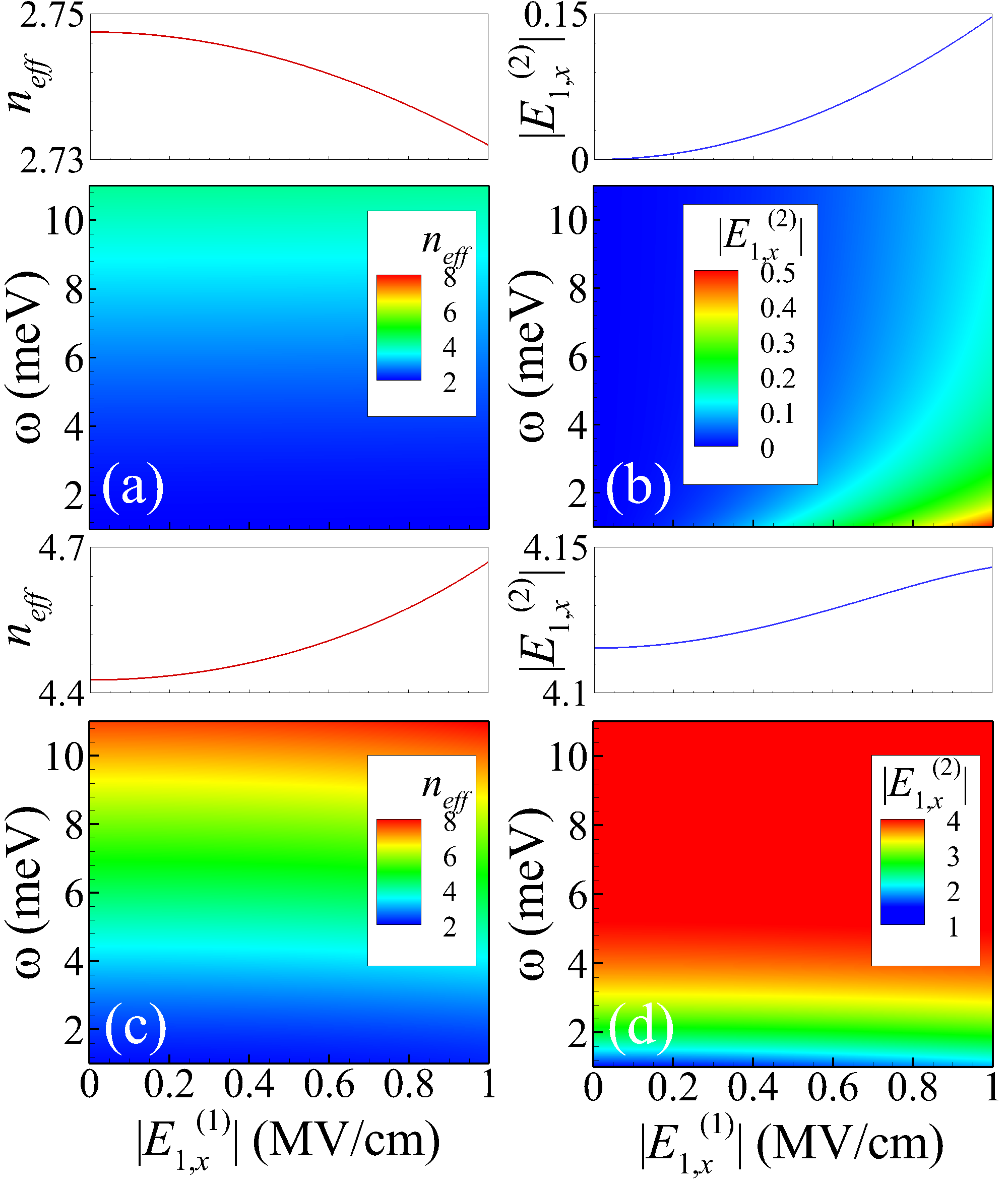}
		\caption{Effective refractive index $n_{eff}$ (a), (c) and second-harmonic in-plane component of electric field $\left|E_{1,x}^{(2)}\right|$ [in MV/cm, (b) and (d)] vs frequency $\omega$ and in-plane component of electric field of first harmonics $\left|E_{1,x}^{(1)}\right|$ of SFSPP [(a), (b)] and DFSPP [(c), (d)]. Upper insets in all panels show the cross section of main panels at fixed frequency $\omega=6\,$meV. All the parameters are the same as those in Fig.\ref{fig:reflectance}(d).}
		\label{fig:eigenmodes}
	\end{figure}
	
	From Fig.~\ref{fig:reflectance}(d), it is evident that the growth of the electric field $E_{1,x}^{(1)}$ results in an increase of the frequency of SFSPP and decrease of the frequency of DFSPP. In more detail, this phenomenon is shown in Fig.~\ref{fig:eigenmodes}, where the effective refractive index $n_{eff}=ck_{x}/\omega$ for SFSPP and DFSPP is depicted by a color map [Figs. \ref{fig:eigenmodes}(a) and \ref{fig:eigenmodes}(c), respectively]. In both cases, the effective refractive index increases with frequency; however, an increase in the electric field $E_{1,x}^{(1)}$ leads to the decrease of the SFSPP effective refractive index [see Fig.~\ref{fig:eigenmodes}(a)] and its increase for DFSPP [Fig.~\ref{fig:eigenmodes}(c)]. From Figs. \ref{fig:eigenmodes}(b) and \ref{fig:eigenmodes}(d), it is evident that an increase of the first-harmonic electric field $E_{1,x}^{(1)}$ is accompanied by the simultaneous increase of the second-harmonic electric field $E_{1,x}^{(2)}$. Nevertheless, if for SFSPP the value of $\left|E_{1,x}^{(1)}\right|$ exceeds $\left|E_{1,x}^{(2)}\right|$ [Fig.~\ref{fig:eigenmodes}(b)], for DFSPP the situation is opposite, here $\left|E_{1,x}^{(2)}\right|\gg\left|E_{1,x}^{(1)}\right|$ [see Fig.~\ref{fig:eigenmodes}(d)]. At the same time, for SFSPP [Fig.~\ref{fig:eigenmodes}(b)] higher frequencies correspond to lower second-harmonic electric field $\left|E_{1,x}^{(2)}\right|$, while for DFSPP [Fig.~\ref{fig:eigenmodes}(d)] the value of $\left|E_{1,x}^{(2)}\right|$ grows with frequency.
	
	\begin{figure}
		\includegraphics[width=8.5cm]{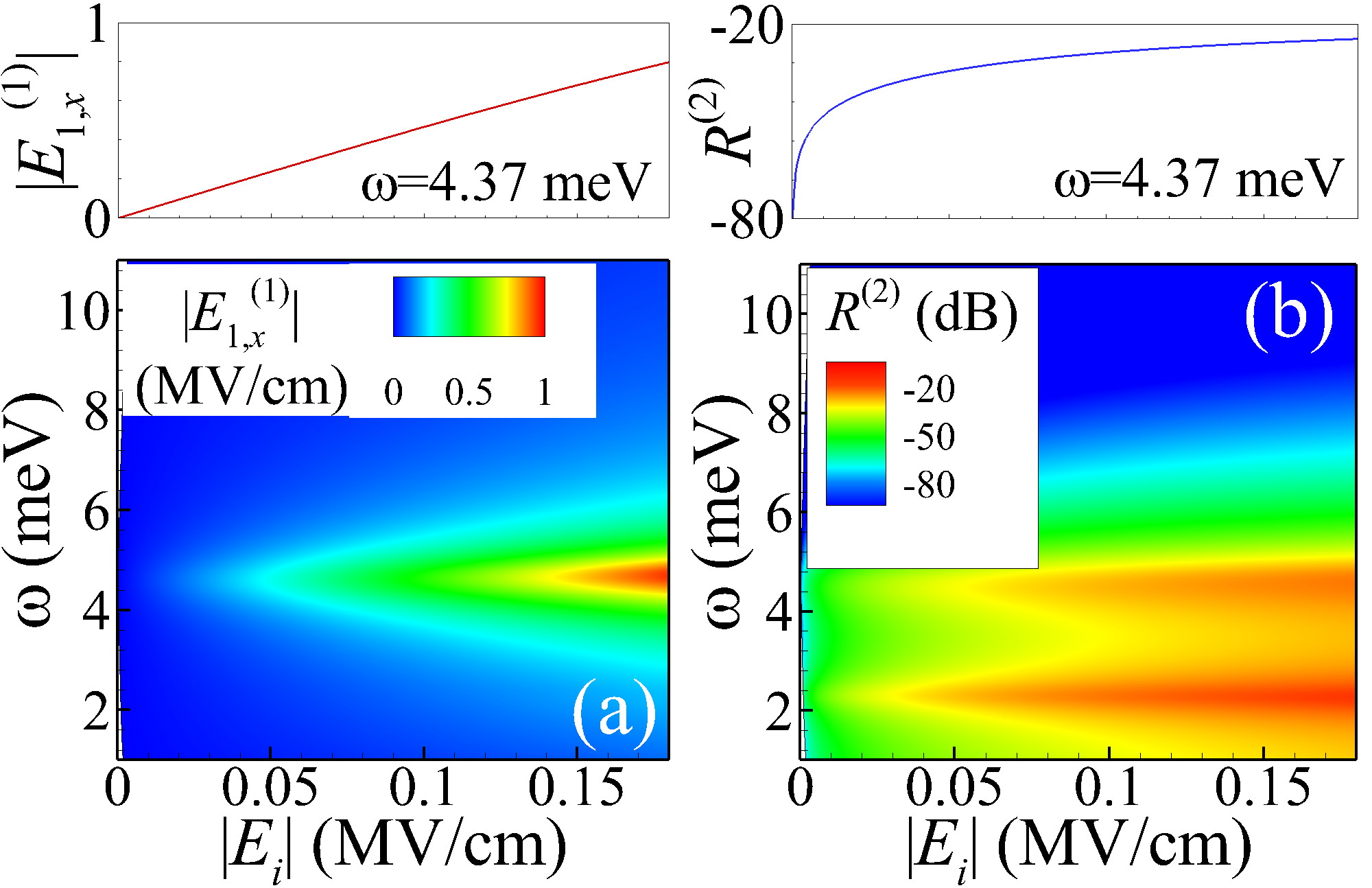}
		\caption{The dependence of the in-plane component of the first-harmonic electric field $\left|E_{1,x}^{(1)}\right|$ (a) and SHG efficiency $R^{(2)}$ (b) on the electric field of the incident wave $\left|E_{i}\right|$ and frequency $\omega$ for the fixed angle of incidence $\theta=40^{\circ}$. Other parameters are the same as those in Figure \ref{fig:reflectance}(a). Upper insets demonstrate the cross sections of dependence in (a) and (b) at fixed frequency $\omega=4.37\,$meV.}
		\label{fig:inc-field}
	\end{figure}
	Since nonlinear phenomena become more pronounced for higher fields, it is natural to presuppose that the SHG efficiency can be enhanced by an increase of the electric field of the incident wave -- this possibility is demonstrated in Fig~\ref{fig:inc-field}. Thus, when the conditions for the resonant excitation of SPP are met [e.g., by a particular choice of angle of incidence and frequency, as in Fig.~\ref{fig:inc-field}(a)], the first-harmonic electric field $E_{1,x}^{(1)}$ is considerably higher than the electric field of the incident wave $E_{i}$ [approximately five times higher for the parameters of Fig.~\ref{fig:inc-field}(a)]. The scale of Fig.~\ref{fig:inc-field}(a) is restricted by the values of the first-harmonic electric field $\left|E_{1,x}^{(1)}\right|<1\,$MV/cm, since higher electric fields can burn out the graphene. This limiting value is achieved at the incident wave's electric field $E_{i}\approx0.18\,$MV/cm and at the resonant frequency $\omega\approx4.38\,$meV. Under those conditions, the SHG efficiency can achieve high enough values $R^{(2)}\sim-20\,$dB, as shown in Fig.~\ref{fig:inc-field}(b) [notice that SHG efficiency is higher for higher values of the incident wave's electric field $E_{i}$].
	
	\section{Bistability in SHG}
	
\begin{figure*}
		\includegraphics[width=17cm]{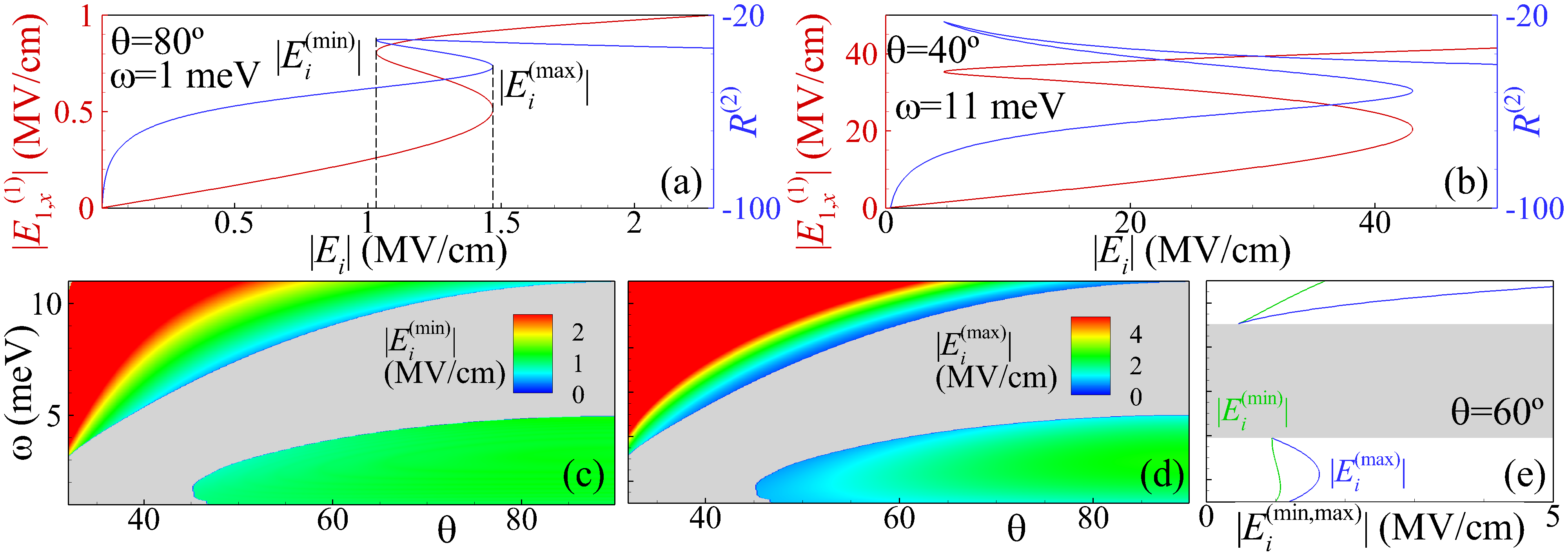}
		\caption{(a),(b) Examples of the bistability of the first-harmonic electric field $E_{1,x}^{(1)}$ (red lines, left $y$ axis) and of SHG efficiency $R^{(2)}$ (blue lines, right $y$ axis) for the parameters of the ATR structure $\theta=80^{0}, \omega=1\,$meV (a); or $\theta=40^{0}, \omega=11\,$meV (b). (c)-(e) Domains of bistability--minimal and $\left|E_{i}^{(min)}\right|$ and maximal $\left|E_{i}^{(max)}\right|$ values of bistability zone vs frequency $\omega$ and angle of incidence $\theta$ [(c) and (d), respectively] or vs frequency $\omega$ for fixed value of angle of incidence $\theta=60^{0}$ [(e), where $\left|E_{i}^{(min)}\right|$ and $\left|E_{i}^{(max)}\right|$ are depicted by green and blue solid lines, correspondingly]. In all panels, other parameters are: $d=10\,\mu$m, $\varepsilon_{3}=16$, $\varepsilon_{2}=1$, $\varepsilon_{1}=3.9$, $E_{F}=0.5\,$eV, $\gamma=0.25\,$meV.}
		\label{fig:bistability}
	\end{figure*}

	The system of equations (\ref{eq:E1x-final}) and (\ref{eq:E2x-final}) that governs SHG can be rewritten in terms of the first-harmonic in-plane component of the electric field $E_{1,x}^{(1)}$ as
	\begin{eqnarray}
		\alpha E_{1,x}^{(1)}+\beta\left|E_{1,x}^{(1)}\right|^{2}E_{1,x}^{(1)}=GE_{i},\label{eq:bistab-complex}
	\end{eqnarray}
	where $\alpha=D^{(1)}\left(d\right)$, $\beta=-Z_{0}^{2}\sigma_{xxx}^{(21)}\sigma_{xxx}^{(11)}/D^{(2)}\left(d\right)$. Without loss of generality, $E_{1,x}^{(1)}$ can be considered as a real value (its complex phase can be made zero by choosing the proper phase of $E_{i}$). In this case, after introducing parameters $X=\left(E_{1,x}^{(1)}\right)^{2}$ and $Y=\left|GE_{i}\right|^{2}$, the condition of equality of absolute values of left and right-hand parts of Eq.~(\ref{eq:bistab-complex}) gives the cubic polynomial function
	\begin{eqnarray}
		X\left(\alpha^{\prime}+\beta^{\prime}X\right)^{2}+X\left(\alpha^{\prime\prime}+\beta^{\prime\prime}X\right)^{2}=Y,\label{eq:bistab-real}
	\end{eqnarray}
	where $\alpha^{\prime},\beta^{\prime}$ and $\alpha^{\prime\prime},\beta^{\prime\prime}$ are real and imaginary parts of respective parameters, such as $\alpha=\alpha^{\prime}+i\alpha^{\prime\prime}$, $\beta=\beta^{\prime}+i\beta^{\prime\prime}$. For a particular choice of parameters, this cubic polynomial function can exhibit two different scenarios of behavior. In the first scenario [see Figs.~\ref{fig:bistability}(a) and \ref{fig:bistability}(b)], there is a domain of the incident-wave electric field $\left|E_{i}^{(\min)}\right|\le\left|E_{i}\right|\le\left|E_{i}^{(\max)}\right|$, where three values of the first-harmonic electric field $E_{1,x}^{(1)}$ (as well as three values of SHG efficiency $R^{(2)}$) correspond to one value of $E_{i}$--this situation is known as bistability, which is typical for graphene \cite{gr-bistability-Peres2014-prb,gr-bistability-Jiang2015-oe,gr-bistability-Jiang2019-ome,gr-bistability-Ahn2017-oe,gr-bistability-Dai2015-oe}. In the second scenario, there is a strict correspondence--one value of $E_{i}$ corresponds to one value of $E_{1,x}^{(1)}$ (like in the upper insets of Fig.~\ref{fig:inc-field}).
	
	Domains of parameters where bistability exists can be predicted from the analysis of Eq.~(\ref{eq:bistab-real}). In the general case, the condition $\partial Y/\partial X=0$ gives two extrema (one minimum and one maximum) of function $Y$ at
	\begin{eqnarray}
		X_{\pm}=\frac{1}{3\left(\beta^{\prime^{2}}+\beta^{\prime\prime^{2}}\right)}\left[-2\left(\alpha^{\prime}\beta^{\prime}+\alpha^{\prime\prime}\beta^{\prime\prime}\right)\right.\label{eq:Xpm}\\
		\left.\pm\sqrt{\left(\alpha^{\prime}\beta^{\prime}+\alpha^{\prime\prime}\beta^{\prime\prime}\right)^{2}-3\left(\alpha^{\prime}\beta^{\prime\prime}-\alpha^{\prime\prime}\beta^{\prime}\right)^{2}}\right].\nonumber
	\end{eqnarray}
	Thus, bistability takes place when $X_{\pm}$ are real and positive. From Eq.~(\ref{eq:Xpm}), it is evident that this happens when the condition
	\begin{eqnarray*}
		-\left(\alpha^{\prime}\beta^{\prime}+\alpha^{\prime\prime}\beta^{\prime\prime}\right)>\sqrt{3}\left|\alpha^{\prime}\beta^{\prime\prime}-\alpha^{\prime\prime}\beta^{\prime}\right|
	\end{eqnarray*}
	is fulfilled. The respective domains of existence and nonexistence of bistability are depicted in Figs.~\ref{fig:bistability}(c) and \ref{fig:bistability}(d) by color and gray regions, respectively. From the comparison of Figs.~\ref{fig:bistability}(c) and \ref{fig:bistability}(d) with Fig.~\ref{fig:reflectance}(b), it follows that bistability regions lie beyond both SFSPP and DFSPP resonances. In the general case, regions of bistability consist of two domains--low-frequency and high-frequency ones. In the low-frequency bistability domain, with an increase of frequency, both $\left|E_{i}^{(\min)}\right|$ [Fig.~\ref{fig:bistability}(c)] and $\left|E_{i}^{(\max)}\right|$ [Fig.~\ref{fig:bistability}(d)] attain some local maxima [in more detail, this phenomenon is shown in Fig.~\ref{fig:bistability}(e)], while in the high-frequency domain, $\left|E_{i}^{(\min)}\right|$ and $\left|E_{i}^{(\max)}\right|$ increase monotonically with frequency.
	
	\section{Conclusions}
	
	To conclude, we considered theoretically SHG in the ATR scheme, consisting of the dielectric prism with graphene below it. We demonstrated that when conditions for the excitation of SPPs in the ATR scheme are met, the polariton-assisted SHG becomes considerably more efficient than in the nonresonant case. It is shown that the frequency dependence of SHG efficiency exhibits two peaks. The physical reason for this phenomenon is either resonant enhancement of the first harmonic owing to SPP excitation (with consequent more efficient SHG), or direct resonant enhancement of the second harmonics. We show that the double-resonance model can be described by the coupled-mode SPP equations. It is also demonstrated that the described ATR scheme with SHG can exhibit bistability at certain domains of parameters. The main results and theory formalism presented here can also be applied to other thin plasmonic systems that are characterized by two-dimensional optical conductivities, such as ultrathin noble metals, where plasmon-driven SHG can occur in the mid-infrared to visible spectral range~\cite{rodriguezecharri2021nonlinear}.
	
	\section*{Acknowledgement}
	The work of J.M.A.P., S.S.C., Y.V.B., J.V.L., and N.M.R.P. was supported by the Fundação para a Ciência e Tecnologia (FCT, Portugal) in the framework of the Strategic Funding UID/04650/2025. V.V.K. was supported by National funds, under the Unit CFTC - Centro de F\'{i}sica Te\'{o}rica e Computacional, Reference No. UID/00618/2025, financing period 2025-2029. J.D.C. acknowledges support from Independent Research Fund Denmark (Grant No. 0165-00051B). The Center for Polariton-driven Light-Matter Interactions (POLIMA) is funded by the Danish National Research Foundation (Project No. DNRF165). J.M.A.P. acknowledges support by Funda\c{c}\~{a}o para a Ci\^{e}ncia e Tecnologia (FCT, Portugal) through Grant. No. 2023.02155.BD. 
	
	\section*{Data Availability}
	The data that support the findings of this article are not publicly available. The data will be available from the author upon reasonable request.
	
	\appendix
	
	\section{Particular solutions of Maxwell equations in different dielectric media}
	
	\label{sec:Maxwell}Maxwell's equations for the TM-polarized wave in medium with dielectric permittivity $\varepsilon_{m}$ ($m=1,...,3$) are written as
	\begin{subequations}
		\begin{equation}
			-\frac{\partial\mathcal{H}_{m,y}}{\partial z}=\frac{\partial\mathcal{D}_{m,x}}{\partial t},\label{eq:MaxEx}
		\end{equation}
		\begin{equation}
			\frac{\partial\mathcal{H}_{m,y}}{\partial x}=\frac{\partial\mathcal{D}_{m,z}}{\partial t},\label{eq:MaxEz}
		\end{equation}
		\begin{equation}
			\frac{\partial\mathcal{E}_{m,x}}{\partial z}-\frac{\partial\mathcal{E}_{m,z}}{\partial x}=-\mu_{0}\frac{\partial\mathcal{H}_{m,y}}{\partial t}.\label{eq:MaxHy}
		\end{equation}
	\end{subequations}
	
	The particular solution of Maxwell's equations (\ref{eq:MaxEx})--(\ref{eq:MaxHy}) inside the prism $\left(m=3\right)$ will have the form
	\begin{align}
		&\left(\begin{array}{c}
			{\cal E}_{3,x}(x,z,t)\\
			Z_{0}{\cal H}_{3,y}(x,z,t)
		\end{array}\right)
		=\sum_{n=1}^{2}\hat{F}_{3}^{(n)}\label{eq:EH3-mat}
		\nonumber\\
		&\times\left(\begin{array}{c}
			\delta_{n,1}E_{i}e^{ink_{3,z}\left(z+d\right)}\\
			E_{3,x}^{(n)}e^{-ink_{3,z}\left(z+d\right)}
		\end{array}\right)e^{in(k_{x}x-\omega t)}+{\rm c.c.} 
	\end{align}
	Hereafter we use notation
	\begin{eqnarray}
		\hat{F}_{m}^{(n)}=\left(\begin{array}{cc}
			1 & 1\\
			\frac{\varepsilon_{m}\omega}{ck_{m,z}} & -\frac{\varepsilon_{m}\omega}{ck_{m,z}}
		\end{array}\right)\label{eq:Fm}
	\end{eqnarray}
	for the field matrix related to medium $m$, $Z_{0}=\sqrt{\mu_{0}/\varepsilon_{0}}$ is the impedance of free space, $k_{m,z}=\sqrt{\omega^{2}\varepsilon_{m}/c^{2}-k_{x}^{2}}$, $k_{x}=\left(\omega/c\right)\sqrt{\varepsilon_{3}}\sin\theta$ are the $z$ and $x$ components of the wavevector, respectively, and $\delta_{n,1}$ is the Kronecker delta. In alternative manner, Eq.\,\eqref{eq:EH3-mat} can be represented as 
	\begin{align}
		&\left(\begin{array}{c}
		{\cal E}_{3,x}(x,z,t)\\
		Z_{0}{\cal H}_{3,y}(x,z,t)
		\end{array}\right)
		= \left(\begin{array}{c}
			E_{i}\\
			Z_{0}H_i
		\end{array}\right)e^{i(k_{x}x+k_{3,z}\left(z+d\right)-\omega t)}+\nonumber\\
		&\sum_{n=1}^{2}\label{eq:EH3-mat-alt}
		\left(\begin{array}{c}
		E_{3,x}^{(n)}\\
		Z_0H_{3,y}^{(n)}
		\end{array}\right)e^{in(k_{x}x-ik_{3,z}\left(z+d\right)-\omega t)}+{\rm c.c.}, 		
	\end{align}
	where $H_i=\sqrt{\varepsilon_{3}}E_i/(Z_0\cos\theta)$, $H_{3,y}^{(n)}=-\sqrt{\varepsilon_{3}}E_{3,x}^{(n)}/(Z_0\cos\theta)$.
	Equations (\ref{eq:EH3-mat}) and (\ref{eq:EH3-mat-alt}) describes the linear superposition of the incident wave with the $x$ component of its electric field amplitude $E_{i}$ (which propagates in the positive direction of the $z$ axis), and reflected first and second harmonics, propagating into the negative $z$-direction, for which the in-plane components of the electric fields are $E_{3,x}^{(n)}$. Here, the presence of the Kronecker delta $\delta_{n,1}$ indicates the absence of an incident second-harmonic wave.
	
	In a similar manner, electromagnetic fields inside the semi-infinite substrate [solution of Maxwell's equations (\ref{eq:MaxEx})--(\ref{eq:MaxHy}) for $m=1$] can be expressed as
	\begin{eqnarray}
		\left(\begin{array}{c}
			{\cal E}_{1,x}(x,z,t)\\
			Z_{0}{\cal H}_{1,y}(x,z,t)
		\end{array}\right)=\sum_{n=1}^{2}\hat{F}_{1}^{(n)}\left(\begin{array}{c}
			E_{1,x}^{(n)}e^{ik_{1,z}z}\\
			0
		\end{array}\right)\label{eq:EH1-mat}
		\nonumber \\
		\times e^{in(k_{x}x-\omega t)}+{\rm c.c.}, 
	\end{eqnarray}
	with the field matrix defined by  
	Eq.~(\ref{eq:Fm}). The zero entry in the second row of 
	Eq.~(\ref{eq:EH1-mat}) means the absence of backward-propagating waves coming from $z=\infty$, thus both harmonics of the electromagnetic field describe forward-propagating transmitted waves. Similar to Eq.\,\eqref{eq:EH3-mat-alt}, the expression for the transmitted wave \eqref{eq:EH1-mat} can be represented as
	\begin{align}
		&\left(\begin{array}{c}
			{\cal E}_{1,x}(x,z,t)\\
			Z_{0}{\cal H}_{1,y}(x,z,t)
		\end{array}\right)
		= \sum_{n=1}^{2}\label{eq:EH1-mat-alt}
		\left(\begin{array}{c}
			E_{1,x}^{(n)}\\
			Z_0H_{1,y}^{(n)}
		\end{array}\right)\nonumber\\
		&\times e^{in(k_{x}x+ik_{1,z}z-\omega t)}+{\rm c.c.}, 		
	\end{align}	
	where $H_{1,y}^{(n)}=\varepsilon_{1}E_{1,x}^{(n)}/\left(Z_{0}\sqrt{\varepsilon_{1}-\varepsilon_{3}\sin^{2}\theta}\right)$.
	
	Inside the semi-infinite media $m=1$ and $m=3$, electromagnetic waves are parametrized by $x-$components of amplitudes of the electric field of forward- and backward-propagating waves. In the medium $m=2$ (which is the medium of finite thickness), the electromagnetic field is expressed in terms of amplitudes of tangential components of fields at the boundary $z=-d$, i.e., $E_{2,x}^{(n)}\left(-d\right)$ and $H_{2,y}^{(n)}\left(-d\right)$, namely
	\begin{eqnarray}
		\label{eq:EH2-mat}
		& &\left(\begin{array}{c}
			{\cal E}_{2,x}(x,z,t)
			\\
			Z_{0}{\cal H}_{2,y}(x,z,t)
		\end{array}\right)
		=\sum_{n=1}^{2}\hat{Q}_{2}^{(n)}(z) 
		\nonumber \\
		& &\times\left(\begin{array}{c}
			E_{2,x}^{(n)}\left(-d\right)
			\\
			Z_{0}H_{2,x}^{(n)}\left(-d\right)
		\end{array}\right)e^{in(k_{x}x-\omega t)}+{\rm c.c.},
		%\nonumber
	\end{eqnarray}
	where
	\begin{align*}
		\hat{Q}_{2}^{(n)}(z)=\left(\begin{array}{cc}
			\cos\left[nk_{2,z}(z+d)\right] & \frac{ick_{2,z}}{\varepsilon_{2}\omega}\sin\left[nk_{2,z}(z+d)\right]\\
			\frac{i\omega\varepsilon_{2}}{ck_{2,z}}\sin\left[nk_{2,z}(z+d)\right] & \cos\left[nk_{2,z}(z+d)\right]
		\end{array}\right)
	\end{align*}
	denotes the transfer matrix.
	
	\section{Calculation of the Linear and Nonlinear Conductivity of Graphene}
	
	\label{sec:Ab-initio-calculation}
	
	\subsection{Solution of Boltzmann equation}
	
	To solve the Boltzmann equation (\ref{eq:Boltzmann}), we consider the electric field as a small perturbation of order $\rho\ll1$, i.e., $\mathcal{E}_{1,x}\left(x,t\right)=\rho\Lambda\left(x,t\right)$, where $\Lambda\left(x,t\right)$ is the scaled electric field. In this case, it is convenient to seek the solution of the Boltzmann equation in the form of a series
	\begin{eqnarray}
		f=\psi_{0}+\rho\psi_{1}+\rho^{2}\psi_{2}\label{eq:expansion-f}
	\end{eqnarray}
	with respect to powers of the small parameter $\rho$ (the zeroth term corresponds to the nonperturbed Fermi-Dirac distribution function $\psi_{0}=f_{0}$). Taking into account that for the Fermi-Dirac distribution $\partial f_{0}/\partial t=\partial f_{0}/\partial x\equiv0$, substituting Eq.~(\ref{eq:expansion-f}) into the Boltzmann equation (\ref{eq:Boltzmann}), and collecting terms in front of different powers of $\rho$, we obtain the recurrence relation
	\begin{eqnarray}
		\frac{\partial\psi_{n}}{\partial t}+v_{x}\frac{\partial\psi_{n}}{\partial x}+\gamma\psi_{n}=\frac{\left|e\right|}{\hbar}\Lambda\left(x,t\right)\frac{\partial\psi_{n-1}}{\partial q_{x}}.\label{eq:recurrence}
	\end{eqnarray}
	This equation can be solved by means of the Green's function
	\begin{align}
		\psi_{n}\left(x,t,\mathbf{q},\omega\right)=\int_{-\infty}^{\infty}dx^{\prime}\int_{-\infty}^{\infty}dt^{\prime}G\left(x-x^{\prime},t-t^{\prime}\right)\label{eq:recurr-solution}\\
		\times\frac{\left|e\right|}{\hbar}\Lambda\left(x^{\prime},t^{\prime}\right)\frac{\partial\psi_{n-1}\left(x^{\prime},t^{\prime},\mathbf{q},\omega\right)}{\partial q_{x}},\nonumber
	\end{align}
	where the Green's function $G\left(x,t\right)$ is obtained as the solution of the equation
	\begin{eqnarray}
		\frac{\partial G}{\partial t}+v_{x}\frac{\partial G}{\partial x}+\gamma G=\delta\left(t\right)\delta\left(x\right).\label{eq:G}
	\end{eqnarray}
	Here, $\delta\left(\cdot\right)$ is the Dirac delta. Defining the Fourier transform of the Green's function as
	\begin{eqnarray*}
		g\left(\kappa,\omega\right)=\frac{1}{2\pi}\int_{-\infty}^{\infty}dxdtG\left(x,t\right)\exp\left(-i\kappa x+i\omega t\right),
	\end{eqnarray*}
	substituting the inverse Fourier transform
	\begin{eqnarray*}
		G\left(x,t\right)=\frac{1}{2\pi}\int_{-\infty}^{\infty}d\omega d\kappa g\left(\kappa,\omega\right)\exp\left(i\kappa x-i\omega t\right)
	\end{eqnarray*}
	into Eq.~(\ref{eq:G}), and taking into account the inverse Fourier transform of delta functions,
	\begin{eqnarray}
		\delta\left(x\right)=\frac{1}{2\pi}\int_{-\infty}^{\infty}d\kappa\exp\left(i\kappa x\right),\label{eq:Fourier-deltax}\\
		\delta\left(t\right)=\frac{1}{2\pi}\int_{-\infty}^{\infty}d\omega\exp\left(-i\omega t\right),\label{eq:Fourier-deltat}
	\end{eqnarray}
	we obtain an expression for the Fourier transform of the Green's function
	\begin{eqnarray*}
		g\left(\kappa,\omega\right)=\frac{1}{2\pi}\frac{1}{\gamma-i\omega+iv_{x}\kappa},
	\end{eqnarray*}
	and, as a result
	\begin{eqnarray}
		G\left(x,t\right)=\frac{1}{\left(2\pi\right)^{2}}\int_{-\infty}^{\infty}d\omega d\kappa\frac{\exp\left(i\kappa x-i\omega t\right)}{\gamma-i\omega+iv_{x}\kappa}.\label{eq:G-integral}
	\end{eqnarray}
	\begin{figure}
		\includegraphics[width=8.5cm]{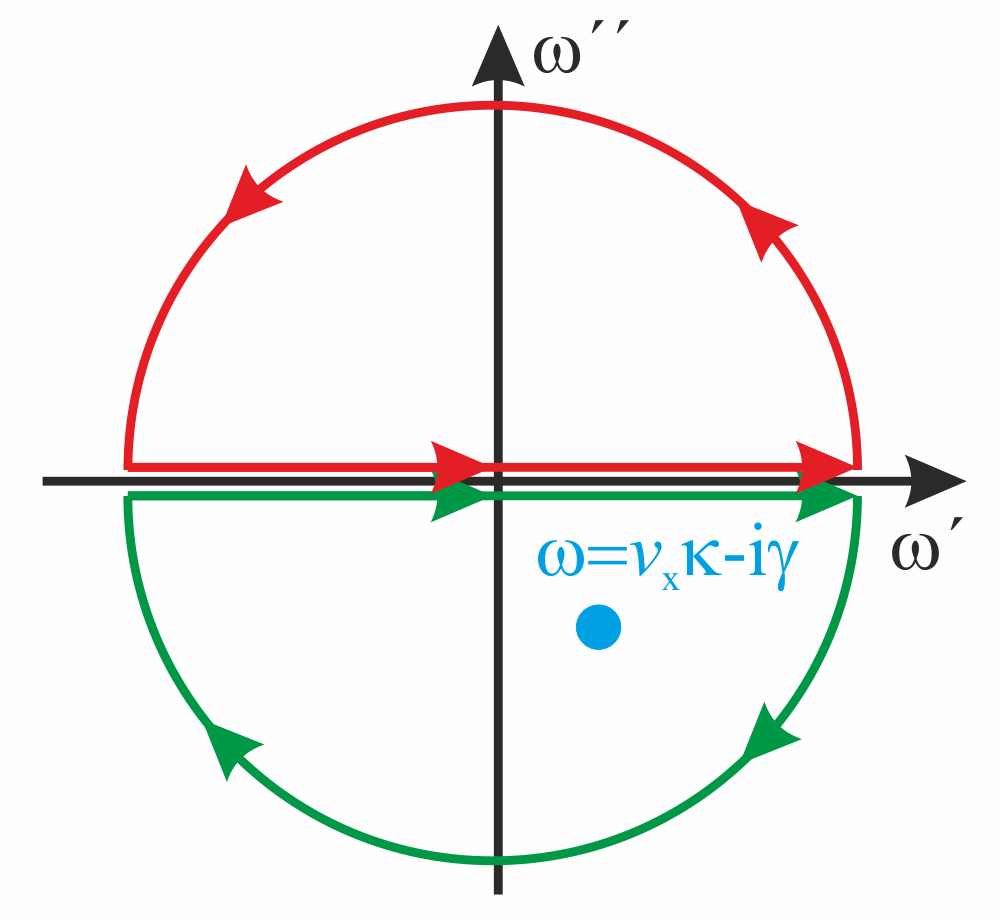}
		\caption{Integration contours for Eq.~(\ref{eq:G-integral}) in the complex plane of frequency $\omega=\omega^{\prime}+i\omega^{\prime\prime}$.}
		\label{fig:contours}
	\end{figure}
	An integral in the above expression is calculated separately for the cases $t<0$ and $t>0$. For the first case, $t<0$, the integration contour should be closed in the upper half-plane of the complex frequency plane (red contour in Fig.\,\ref{fig:contours}). In this situation, the integral along the arc in the upper half-plane is zero, the pole $\omega=v_{x}\kappa-i\gamma$ of the integral in Eq.~(\ref{eq:G-integral}) is located beyond the integration contour, and, as a result
	\begin{eqnarray}
		G\left(x,t\right)\equiv0.\label{eq:G0}
	\end{eqnarray}
	For the second case, $t>0$, the integration contour should be closed in the lower half-plane (green contour in Fig.\,\ref{fig:contours}). In this situation, the integral along the arc is also zero, but the pole is located inside the integration contour, thus Eq.~(\ref{eq:G-integral}) can be represented as
	\begin{widetext}
		\begin{eqnarray}
			G\left(x,t\right)=-2\pi i\mathrm{Res}\left.\left[\frac{i}{\left(2\pi\right)^{2}}\int_{-\infty}^{\infty}d\kappa\frac{\exp\left(i\kappa x-i\omega t\right)}{\omega-v_{x}\kappa+i\gamma}\right]\right|_{\omega=v_{x}\kappa-i\gamma}.\label{eq:G-res}
		\end{eqnarray}
		Since the pole in Eq.~(\ref{eq:G-res}) is simple, taking into account the Fourier transform of the delta function (\ref{eq:Fourier-deltax}), we obtain
		\begin{eqnarray*}
			G\left(x,t\right)=\exp\left[-\gamma t\right]\delta\left(x-v_{x}t\right).
		\end{eqnarray*}
		This expression, combined with Eq.~(\ref{eq:G0}), gives the general form of the retarded Green's function,
		\begin{eqnarray*}
			G\left(x-x^{\prime},t-t^{\prime}\right)=\left\{\begin{array}{cc}
				0, & \qquad t^{\prime}>t\\
				\exp\left[-\gamma\left(t-t^{\prime}\right)\right]\delta\left(x-x^{\prime}-v_{x}\left(t-t^{\prime}\right)\right), & \qquad t^{\prime}<t
			\end{array}\right..
		\end{eqnarray*}
		This Green's function, being substituted into Eq.~(\ref{eq:recurr-solution}), for $f_{n}=\rho^{n}\psi_{n}$ gives Eq.~(\ref{eq:ite_solution_f}).
		
		Further, substituting the electric field [first line in Eq.~(\ref{eq:EH1-mat})] as well as the derivative (\ref{eq:df0-dqx}) into Eq.~(\ref{eq:ite_solution_f}), we obtain an expression for the first-order correction in the form
		\begin{eqnarray*}
			f_{1}\left(x,t,\mathbf{q},\omega\right)=-\frac{\left|e\right|}{\hbar}\delta\left(q-q_{F}\right)\frac{q_{x}}{q}\exp\left[-\gamma t\right]\sum_{n=1}^{2}E_{1,x}^{(n)}\exp\left[ink_{x}\left(x-v_{x}t\right)\right]\int_{-\infty}^{t}dt^{\prime}\exp\left[\gamma t^{\prime}+ink_{x}v_{x}t^{\prime}-in\omega t^{\prime}\right],
		\end{eqnarray*}
		which after integration gives Eq.~(\ref{eq:f1_exact_sol}). Differentiating Eq.~(\ref{eq:f1-final}), we obtain
		\begin{eqnarray}
			\frac{\partial f_{1}\left(x,t,\mathbf{q},\omega\right)}{\partial q_{x}}=-\frac{\left|e\right|}{\hbar}\left\{\frac{d\delta\left(q-q_{F}\right)}{dq}\frac{q_{x}^{2}}{q^{2}}+\delta\left(q-q_{F}\right)\frac{q_{y}^{2}}{q^{3}}\right\}\times\nonumber\\
			\times\left\{\sum_{n=1}^{2}E_{1,x}^{\left(n\right)}\frac{\exp\left[ink_{x}x-in\omega t\right]}{\gamma-in\omega}\left(1-\frac{ink_{x}v_{F}}{\gamma-in\omega}\frac{q_{x}}{q}\right)+{\rm c.c.}\right\}+\label{eq:df1-dqx-detail}\\
			+\frac{\left|e\right|}{\hbar}\delta\left(q-q_{F}\right)\frac{q_{x}}{q}\left\{\sum_{n=1}^{2}E_{1,x}^{\left(n\right)}\frac{\exp\left[ink_{x}x-in\omega t\right]ink_{x}v_{F}}{\left(\gamma-in\omega\right)^{2}}\frac{q_{y}^{2}}{q^{3}}+{\rm c.c.}\right\}.\nonumber
		\end{eqnarray}
		Here, we take into account that 
		\begin{eqnarray*}
			\frac{\partial}{\partial q_{x}}\left(\frac{q_{x}}{q}\right)=\frac{\partial}{\partial q_{x}}\left(\frac{q_{x}}{\sqrt{q_{x}^{2}+q_{y}^{2}}}\right)=\frac{1}{\sqrt{q_{x}^{2}+q_{y}^{2}}}-\frac{q_{x}^{2}}{\left(q_{x}^{2}+q_{y}^{2}\right)^{3/2}}=\frac{q_{y}^{2}}{q^{3}}.
		\end{eqnarray*}
		With collecting terms in front of amplitudes $E_{1,x}^{\left(n\right)}$, Eq.~(\ref{eq:df1-dqx-detail}) can be written in a more compact form--see Eq.~(\ref{eq:df1-dqx}). Substituting Eq.~(\ref{eq:df1-dqx}) into Eq.~(\ref{eq:ite_solution_f}), we have
		\begin{eqnarray*}
			f_{2}\left(x,t,\mathbf{q},\omega\right)=-\frac{\left|e\right|^{2}}{\hbar^{2}}\int_{-\infty}^{t}dt^{\prime}\exp\left[-\gamma\left(t-t^{\prime}\right)\right]\left\{\sum_{n=1}^{2}E_{1,x}^{(n)}\exp\left[ink_{x}\left(x-v_{x}\left(t-t^{\prime}\right)\right)-in\omega t^{\prime}\right]+{\rm c.c.}\right\}\\
			\times\left\{\sum_{n^{\prime}=1}^{2}E_{1,x}^{(n^{\prime})}\Gamma_{n^{\prime}}\left(\mathbf{q}\right)\exp\left[in^{\prime}k_{x}\left(x-v_{x}\left(t-t^{\prime}\right)\right)-in^{\prime}\omega t^{\prime}\right]+{\rm c.c.}\right\}. 
		\end{eqnarray*}
		 
		Neglecting harmonics higher than the second ones, the above equation can be rewritten as
		\begin{eqnarray*}
			f_{2}\left(x,t,\mathbf{q},\omega\right)=-\frac{\left|e\right|^{2}}{\hbar^{2}}\exp\left[-\gamma t\right]\int_{-\infty}^{t}dt^{\prime}\exp\left[\gamma t^{\prime}\right]\left\{\left|E_{1,x}^{(1)}\right|^{2}\Gamma_{1}\left(\mathbf{q}\right)+\left|E_{1,x}^{(2)}\right|^{2}\Gamma_{2}\left(\mathbf{q}\right)\right.\\
			+\exp\left[ik_{x}\left(x-v_{x}\left(t-t^{\prime}\right)\right)-i\omega t^{\prime}\right]E_{1,x}^{(2)}\overline{E_{1,x}^{(1)}}\left(\overline{\Gamma_{1}\left(\mathbf{q}\right)}+\Gamma_{2}\left(\mathbf{q}\right)\right)\\
			\left.+\exp\left[i2k_{x}\left(x-v_{x}\left(t-t^{\prime}\right)\right)-i2\omega t^{\prime}\right]\left(E_{1,x}^{(1)}\right)^{2}\Gamma_{1}\left(\mathbf{q}\right)+{\rm c.c.}\right\}.
		\end{eqnarray*}
		This expression, after being integrated and expanded into a Taylor series, transforms into Eq.~(\ref{eq:f2}).
		\end{widetext}
	
	\subsection{Linear and nonlinear currents}
	
	In order to calculate the linear current, we perform the integration in the momentum plane of Eq.~(\ref{eq:jl}). In this case, it is convenient to integrate in polar coordinates
	\begin{eqnarray*}
		q_{x}=q\cos\left(\varphi_{q}\right),\quad q_{x}=q\sin\left(\varphi_{q}\right).
	\end{eqnarray*}
	Thus, substituting Eq.~(\ref{eq:f1-final}) into Eq.~(\ref{eq:jl}), and representing the result in polar coordinates, we have
	\begin{align}
		\label{eq:jl-integral-2}
		\mathcal{J}_{x}^{(L)}\left(x,t\right)=\frac{4\left|e\right|^{2}}{\left(2\pi\right)^{2}\hbar}\intop_{0}^{\infty}qdq\intop_{0}^{2\pi}d\varphi_{q} 
		\nonumber \\
		\times v_{F}\cos^{2}\left(\varphi_{q}\right)\sum_{n=1}^{2}E_{1,x}^{\left(n\right)}\frac{\exp\left[in(k_{x}x-\omega t)\right]}{\gamma-in\omega}
		\nonumber\\
		\times\left(1-\frac{ink_{x}v_{F}\cos\left(\varphi_{q}\right)}{\gamma-in\omega}\right)\delta\left(q-q_{F}\right)+{\rm c.c.}
	\end{align}
	Using the properties of the delta function, the above equation can be represented as 
	\begin{widetext}
		\begin{eqnarray}
			\mathcal{J}_{x}^{(L)}\left(x,t\right)=\frac{4\left|e\right|^{2}v_{F}q_{F}}{\left(2\pi\right)^{2}\hbar}\sum_{n=1}^{2}E_{1,x}^{\left(n\right)}\frac{\exp\left[ink_{x}x-in\omega t\right]}{\gamma-in\omega}\left(\intop_{0}^{2\pi}\cos^{2}\left(\varphi_{q}\right)d\varphi_{q}-\frac{ink_{x}v_{F}}{\gamma-in\omega}\intop_{0}^{2\pi}\cos^{3}\left(\varphi_{q}\right)d\varphi_{q}\right)+{\rm c.c.}\label{eq:jl-integral-1}
		\end{eqnarray}
		In Eq.~(\ref{eq:jl-integral-1}), the second integral is zero, so spatial dispersion (dependence of $k_{x}$) appears only as a correction of order $k_{x}^{2}$ and higher. At the same time, integration of the first integral in Eq.~(\ref{eq:jl-integral-1}) gives the final result for the linear current (\ref{eq:j-l}).
		
		For the nonlinear current, substitution of Eq.~(\ref{eq:f2}) into Eq.~(\ref{eq:j-nl}), and transforming the result into polar coordinates, we obtain
		\begin{eqnarray}
			\label{eq:j-nl-prelim}
			\mathcal{J}_{x}^{(NL)}\left(x,t\right)=\frac{4\left|e\right|^{3}v_{F}}{\left(2\pi\right)^{2}\hbar^{2}}\intop_{0}^{\infty}qdq\intop_{0}^{2\pi}d\varphi_{q}\cos\left(\varphi_{q}\right)\sum_{n=1}^{2}K^{(n)}\frac{\exp\left[in(k_{x}x-\omega t)\right]}{\gamma-in\omega}\left(1-\frac{ink_{x}v_{F}}{\gamma-in\omega}\cos\left(\varphi_{q}\right)\right). 
		\end{eqnarray}
		Introducing the notation for integrals included into Eq.~(\ref{eq:j-nl-prelim}),
		\begin{eqnarray}
			\eta_{ln}=\int_{0}^{\infty}qdq\int_{0}^{2\pi}d\varphi_{q}\cos\varphi_{q}\Gamma_{l}\left(1-\frac{ink_{x}v_{F}\cos\varphi_{q}}{\gamma-in\omega}\right)\label{eq:eta}\\
			\approx\int_{0}^{\infty}qdq\int_{0}^{2\pi}d\varphi\cos\varphi_{q}\left[\frac{\Upsilon\left(\mathbf{q}\right)}{\gamma-il\omega}-ilk_{x}v_{F}\frac{\Phi\left(\mathbf{q}\right)}{\left(\gamma-il\omega\right)^{2}}-\frac{\Upsilon\left(\mathbf{q}\right)}{\gamma-il\omega}\frac{ink_{x}v_{F}\cos\varphi_{q}}{\gamma-in\omega}\right],\nonumber\\
			\zeta_{ln}=\int_{0}^{\infty}qdq\int_{0}^{2\pi}d\varphi_{q}\cos\varphi_{q}\overline{\Gamma_{l}}\left(1-\frac{ink_{x}v_{F}\cos\varphi_{q}}{\gamma-in\omega}\right)\label{eq:zeta}\\
			\approx\int_{0}^{\infty}qdq\int_{0}^{2\pi}d\varphi\cos\varphi_{q}\left[\frac{\Upsilon\left(\mathbf{q}\right)}{\gamma+il\omega}+ilk_{x}v_{F}\frac{\Phi\left(\mathbf{q}\right)}{\left(\gamma+il\omega\right)^{2}}-\frac{\Upsilon\left(\mathbf{q}\right)}{\gamma+il\omega}\frac{ink_{x}v_{F}\cos\varphi_{q}}{\gamma-in\omega}\right],\nonumber
		\end{eqnarray}
		we can represent the nonlinear current as
		\begin{eqnarray}
			\mathcal{J}_{x}^{(NL)}\left(x,t\right)=\frac{\left|e\right|^{3}v_{F}}{\left(\pi\hbar\right)^{2}}\left\{\frac{\left|E_{1,x}^{(1)}\right|^{2}}{\gamma}\eta_{10}+\frac{\left|E_{1,x}^{(2)}\right|^{2}}{\gamma}\eta_{20}\right.\nonumber\\
			\left.+ E_{1,x}^{(2)}\overline{E_{1,x}^{(1)}}\left(\zeta_{11}+\eta_{21}\right)\frac{\exp\left[ik_{x}x-i\omega t\right]}{\gamma-i\omega}+ \left(E_{1,x}^{(1)}\right)^{2}\eta_{12}\frac{\exp\left[i2k_{x}x-i2\omega t\right]}{\gamma-i2\omega}\right\}+{\rm c.c.}\label{eq:j-nl-eta-zeta}
		\end{eqnarray}
		Taking into account that
		\begin{eqnarray*}
			\int_{0}^{\infty}qdq\int_{0}^{2\pi}d\varphi\cos\varphi_{q}\Upsilon\left(\mathbf{q}\right)=\int_{0}^{\infty}dq\int_{0}^{2\pi}d\varphi\cos\varphi_{q}\left\{\sin^{2}\varphi_{q}\delta\left(q-q_{F}\right)+q\cos^{2}\varphi_{q}\frac{\partial\delta\left(q-q_{F}\right)}{\partial q}\right\}=0,\\
			\int_{0}^{\infty}qdq\int_{0}^{2\pi}d\varphi\cos\varphi_{q}\Phi\left(\mathbf{q}\right)=\int_{0}^{\infty}dq\int_{0}^{2\pi}d\varphi_{q}\cos^{2}\varphi_{q}\left\{2\sin^{2}\varphi_{q}\delta\left(q-q_{F}\right)+q\cos^{2}\varphi_{q}\frac{\partial\delta\left(q-q_{F}\right)}{\partial q}\right\}\\
			=\int_{0}^{\infty}dq\int_{0}^{2\pi}d\varphi_{q}\cos^{2}\varphi_{q}\left\{2\sin^{2}\varphi_{q}-\cos^{2}\varphi_{q}\right\}\delta\left(q-q_{F}\right)=-\frac{\pi}{4},\\
			\int_{0}^{\infty}qdq\int_{0}^{2\pi}d\varphi\cos^{2}\varphi_{q}\Upsilon\left(\mathbf{q}\right)=\int_{0}^{\infty}dq\int_{0}^{2\pi}d\varphi\cos^{2}\varphi_{q}\left\{\sin^{2}\varphi_{q}\delta\left(q-q_{F}\right)+q\cos^{2}\varphi_{q}\frac{\partial\delta\left(q-q_{F}\right)}{\partial q}\right\}\\
			=\int_{0}^{\infty}dq\int_{0}^{2\pi}d\varphi\cos^{2}\varphi_{q}\left\{\sin^{2}\varphi_{q}-\cos^{2}\varphi_{q}\right\}\delta\left(q-q_{F}\right)=-\frac{\pi}{2},
		\end{eqnarray*}
\end{widetext}
		Eqs.~(\ref{eq:eta}) and (\ref{eq:zeta}) can be represented as
		\begin{eqnarray*}
			\eta_{ln}=\frac{\pi}{4}\frac{ik_{x}v_{F}}{\gamma-il\omega}\left[\frac{l}{\gamma-il\omega}+\frac{2n}{\gamma-in\omega}\right],\\
			\zeta_{ln}=-\frac{\pi}{4}\frac{ik_{x}v_{F}}{\gamma+il\omega}\left[\frac{l}{\gamma+il\omega}-\frac{2n}{\gamma-in\omega}\right].
		\end{eqnarray*}
		Substituting these expressions into Eq.~(\ref{eq:j-nl-eta-zeta}) and neglecting the zeroth harmonics, it is possible to obtain the expression for the nonlinear current in the form of Eq.~(\ref{eq:j-nl-final}).

	\section{Dependence of SHG Efficiency on the Parameters of Structure}
	
\begin{figure*}
		\includegraphics[width=17cm]{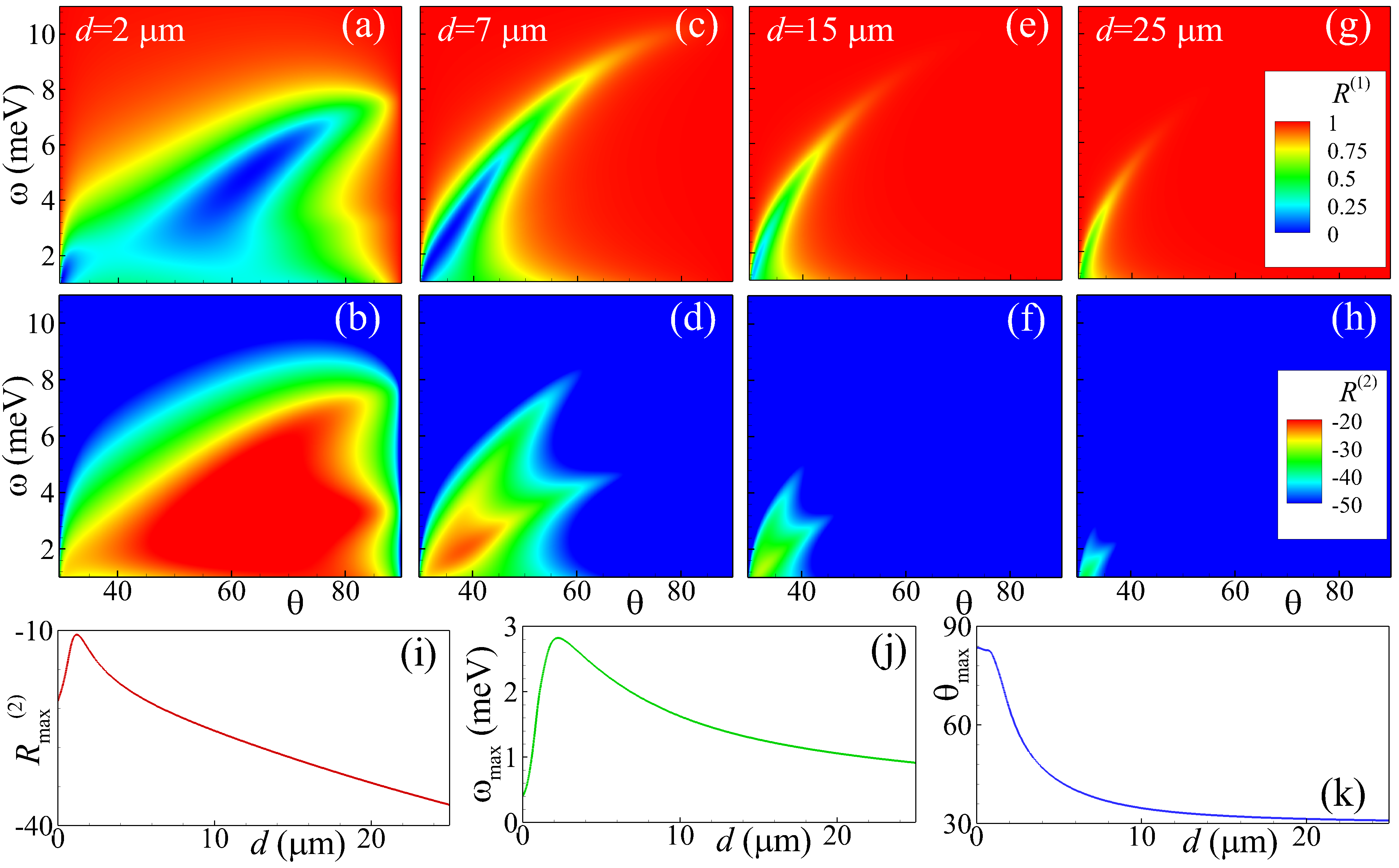}
		\caption{(a)--(h) Reflectance $R^{(1)}$ [top row, (a), (c), (e), and (g)] and SHG efficiency $R^{(2)}$ [middle row, (b), (d), (f), and (h)] vs angle of incidence $\theta$ and frequency $\omega$ of ATR structure for different values of distance between prism and graphene $d=2\,\mu$m (a), (b), $d=7\,\mu$m [(c), (d)], $d=15\,\mu$m (e),  (f), $d=25\,\mu$m (g), (h). (i)--(k) Dependence of the maximal SHG efficiency $R_{\rm max}^{(2)}$ (i), frequency $\omega_{\rm max}$ (j), and angle of incidence $\theta_{\rm max}$ (k) on distance between prism and graphene $d$. In all panels, other parameters are $d=10\,\mu$m, $\varepsilon_{3}=16$, $\varepsilon_{2}=1$, $\varepsilon_{1}=3.9$, $E_{F}=0.5\,$eV, $\gamma=0.5\,$meV, $E_{i}=0.1\,$MV/cm.}
		\label{fig:R12-D}
	\end{figure*}

	In order to define the parameters that provide maximal efficiency of the SHG, we investigate its dependence on the parameters of the ATR structure. The dependence of SHG efficiency on the distance between the prism and graphene $d$ is demonstrated in Fig.~\ref{fig:R12-D}. From the comparison of Figs.~\ref{fig:R12-D}(a), \ref{fig:R12-D}(c), \ref{fig:R12-D}(e), and \ref{fig:R12-D}(g), it can be seen that an increase in distance $d$ results in a smaller depth of reflectance $R^{(1)}$--the excitation of SPPs turns out to be less pronounced in the ATR scheme. Simultaneously, an increase in distance $d$ in the range $d>2\,\mu$m leads to the decrease of SHG efficiency $R^{(2)}$ [compare Figures \ref{fig:R12-D}(b), \ref{fig:R12-D}(d), \ref{fig:R12-D}(f), and \ref{fig:R12-D}(h)]. The dependence of maximal values of SHG efficiency $R^{(2)}_{max}$ as well as the frequency $\omega_{max}$ and angle of incidence $\theta_{max}$ at which these maximal SHG efficiencies take place are shown in Figs.~\ref{fig:R12-D}(i)--\ref{fig:R12-D}(k), respectively, as functions of the prism-graphene distance $d$. Thus, the maximal SHG efficiency ($\sim-11\,$dB) is achieved at a prism-graphene distance $d\approx 1.25\,\mu$m [see Fig.~\ref{fig:R12-D}(i)] and decreases monotonically at larger values of $d$. These maximal values of SHG are achieved in the frequency range $0.5\,$meV $<\omega_{max}\lesssim 3\,$meV [Fig.~\ref{fig:R12-D}(j)], and angles of incidence in the range $30^{\circ}<\theta_{max}\lesssim 85^{\circ}$ [Fig.~\ref{fig:R12-D}(k)].
	
\begin{figure*}
		\includegraphics[width=17cm]{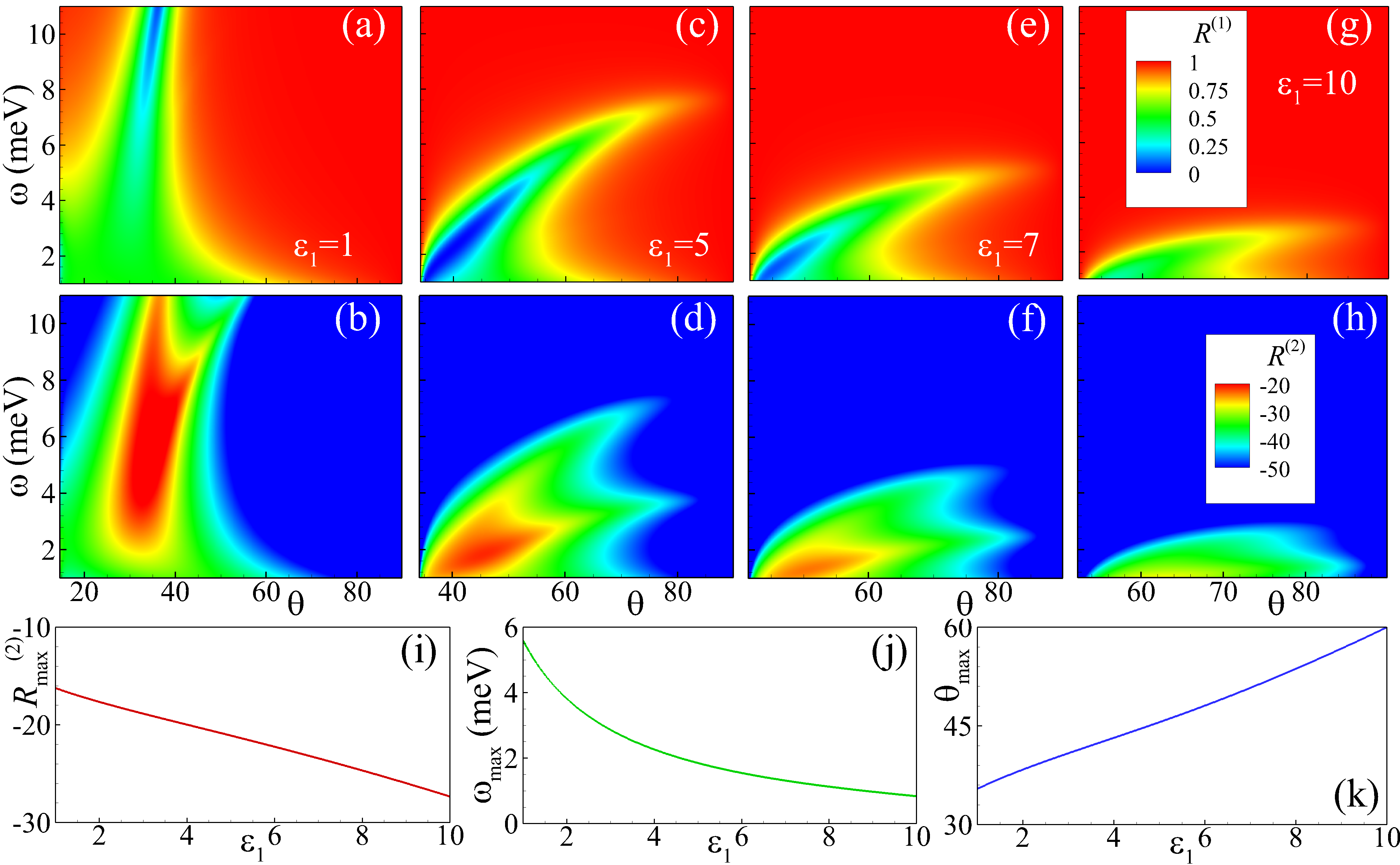}
		\caption{(a)--(h) Reflectance $R^{(1)}$ [top row, (a), (c), (e), and (g)] and SHG efficiency $R^{(2)}$ [middle row, (b), (d), (f), and (h)] vs angle of incidence $\theta$ and frequency $\omega$ of ATR structure for different relative dielectric permittivity of substrate $\varepsilon_1=1$ (a), (b), $\varepsilon_1=5$ (c), (d), $\varepsilon_1=7$ (e), (f), $\varepsilon_1=10$ (g), (h). (i)--(k) Dependence of the maximal SHG efficiency $R_{\rm max}^{(2)}$ (i), frequency $\omega_{\rm max}$ (j), and angle of incidence $\theta_{\rm max}$ (k) on permittivity of substrate $\varepsilon_1$. In all panels, other parameters are $d=5\,\mu$m, $\varepsilon_{3}=16$, $\varepsilon_{2}=1$, $E_{F}=0.5\,$eV, $\gamma=0.5\,$meV, $E_{i}=0.1\,$MV/cm.}
		\label{fig:R12-EpsS}
	\end{figure*}

	The dependence of SHG efficiency on the relative dielectric permittivity of the substrate $\varepsilon_1$, shown in Fig.~\ref{fig:R12-EpsS}, demonstrates both the decrease of SHG efficiency and a smaller depth of reflectance $R^{(1)}$ with the growth of the substrate's permittivity $\varepsilon_1$ [compare Figs.~\ref{fig:R12-EpsS}(b), \ref{fig:R12-EpsS}(d), \ref{fig:R12-EpsS}(f), and \ref{fig:R12-EpsS}(h) as well as Figs.~\ref{fig:R12-EpsS}(a), \ref{fig:R12-EpsS}(c), \ref{fig:R12-EpsS}(e), and \ref{fig:R12-EpsS}(g)]. In more detail, this phenomenon is shown in Fig.~\ref{fig:R12-EpsS}(i), where the maximal SHG effectiveness drops with an increase of $\varepsilon_1$ [as well as the frequency $\omega_{max}$ at which the maximal SHG efficiency takes place, see Fig.~\ref{fig:R12-EpsS}(j)]. At the same time, the angle of incidence corresponding to the maximal SHG efficiency $\theta_{max}$ is a monotonically increasing function of the substrate's permittivity [Fig.~\ref{fig:R12-EpsS}(k)]. As a result, the best SHG efficiency should occur for the case of suspended graphene, where $\varepsilon_1=1$.
	
\begin{figure*}
		\includegraphics[width=17cm]{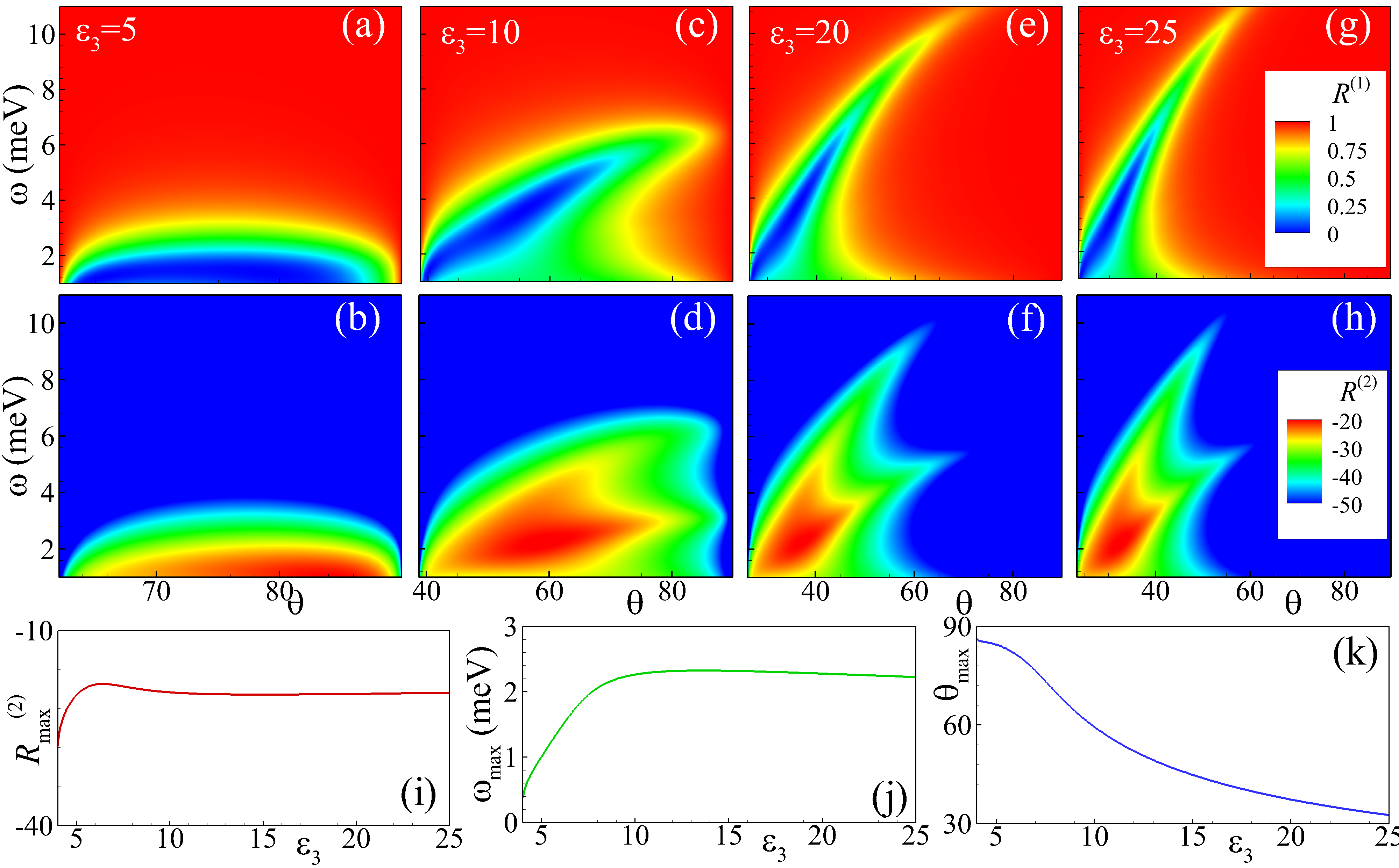}
		\caption{(a)--(h) Reflectance $R^{(1)}$ [top row, (a), (c), (e), and (g)] and SHG efficiency $R^{(2)}$ [middle row, (b), (d), (f), and (h)] vs angle of incidence $\theta$ and frequency $\omega$ of ATR structure for different dielectric permittivity of prism $\varepsilon_3=5$ (a), (b), $\varepsilon_3=10$ (c), (d), $\varepsilon_3=20$ (e), (f), $\varepsilon_3=25$ (g), (h). (i)--(k) Dependence of the maximal SHG efficiency $R_{\rm max}^{(2)}$ (i), frequency $\omega_{\rm max}$ (j), and angle of incidence $\theta_{\rm max}$ (k) on dielectric permittivity of prism $\varepsilon_3$. In all panels, other parameters are $d=5\,\mu$m, $\varepsilon_{2}=1$, $\varepsilon_{1}=3.9$, $E_{F}=0.5\,$eV, $\gamma=0.5\,$meV, $E_{i}=0.1\,$MV/cm.}
		\label{fig:R12-EpsP}
	\end{figure*}

	The dependence of reflectance $R^{(1)}$ on the dielectric permittivity of the prism $\varepsilon_{3}$ demonstrates an increase of the sharpness of the resonance with an increase of $\varepsilon_{3}$ [as it follows from the comparison of Figs.~\ref{fig:R12-EpsP}(a), \ref{fig:R12-EpsP}(c), \ref{fig:R12-EpsP}(e), and \ref{fig:R12-EpsP}(g)]. At the same time, the SHG effect turns out to be more pronounced for higher values of $\varepsilon_3$ [Figs.~\ref{fig:R12-EpsP}(b), \ref{fig:R12-EpsP}(d), \ref{fig:R12-EpsP}(f), and \ref{fig:R12-EpsP}(h)]. The maximal value of SHG efficiency [Fig.~\ref{fig:R12-EpsP}(i)] after achieving the maximal value at $\varepsilon_{3}\approx 6$ turns out to be almost not sensitive to $\varepsilon_{3}$ [similar to the frequency $\omega_{max}$, see Fig.~\ref{fig:R12-EpsP}(j)]. In contrast, the angle of incidence $\theta_{max}$ (at which the maximal value of SHG is achieved) is a decreasing function of $\varepsilon_{3}$ [Fig.~\ref{fig:R12-EpsP}(k)].
	
\begin{figure*}
		\includegraphics[width=17cm]{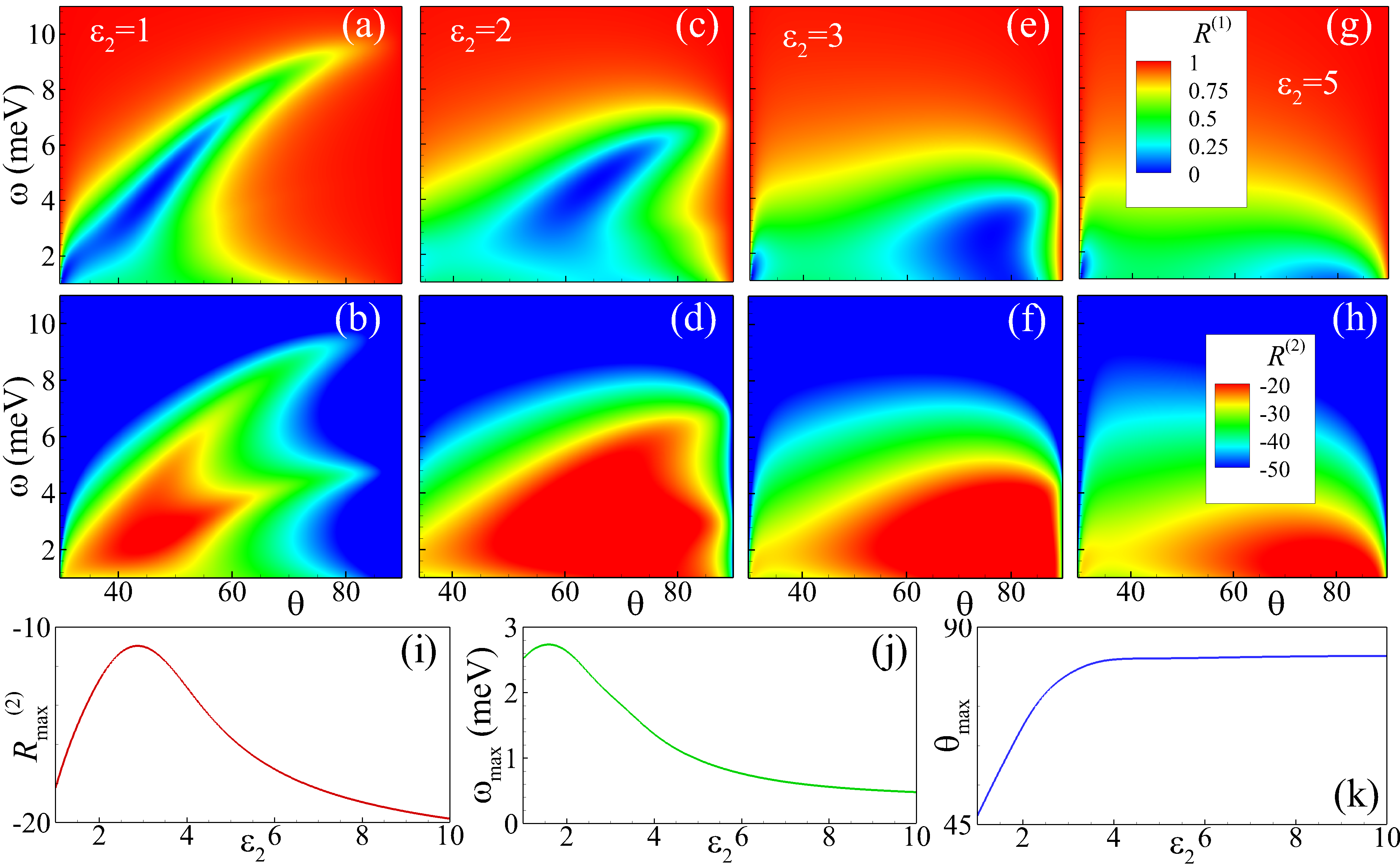}
		\caption{(a)--(h) Reflectance $R^{(1)}$ [top row, (a), (c), (e), and (g)] and SHG efficiency $R^{(2)}$ [middle row, (b), (d), (f), and (h) vs angle of incidence $\theta$ and frequency $\omega$ of ATR structure for different values of the dielectric permittivity $\varepsilon_2=1$ (a), (b), $\varepsilon_2=2$ (c), (d), $\varepsilon_2=3$ (e), (f), $\varepsilon_2=5$ (g), (h). (i)--(k) Dependence of the maximal SHG efficiency $R_{\rm max}^{(2)}$ (i), frequency $\omega_{\rm max}$ (j), and angle of incidence $\theta_{\rm max}$ (k) on dielectric permittivity $\varepsilon_2$. In all panels, other parameters are $d=4\,\mu$m, $\varepsilon_{3}=16$, $\varepsilon_{1}=3.9$, $E_{F}=0.5\,$eV, $\gamma=0.5\,$meV, $E_{i}=0.1\,$MV/cm.}
		\label{fig:R12-EpsT}
	\end{figure*}

	As it follows from the comparison of Figs.~\ref{fig:R12-EpsT}(a), \ref{fig:R12-EpsT}(c), \ref{fig:R12-EpsT}(e), and \ref{fig:R12-EpsT}(g), with an increase of dielectric permittivity of the media between graphene and prism $\varepsilon_{2}$, the ATR resonance becomes less sharp and is shifted to lower frequencies (red-shifted). This phenomenon is accompanied by a decreasing SHG efficiency $R^{(2)}$ [Figs.~\ref{fig:R12-EpsT}(b), \ref{fig:R12-EpsT}(d), \ref{fig:R12-EpsT}(f), and \ref{fig:R12-EpsT}(h)]. Notice that the SHG efficiency achieves its maximal value at $\varepsilon_2\approx 2.9$, as shown in Fig.~\ref{fig:R12-EpsT}(i), while $\omega_{max}$ is maximal at $\varepsilon_2\approx 1.5$ [Fig.~\ref{fig:R12-EpsT}(j)]. The angle of incidence is a monotonically increasing function of $\varepsilon_2$ and it saturates in the interval $\varepsilon_2>3$ [Fig.~\ref{fig:R12-EpsT}(k)].
	
\begin{figure*}
		\includegraphics[width=17cm]{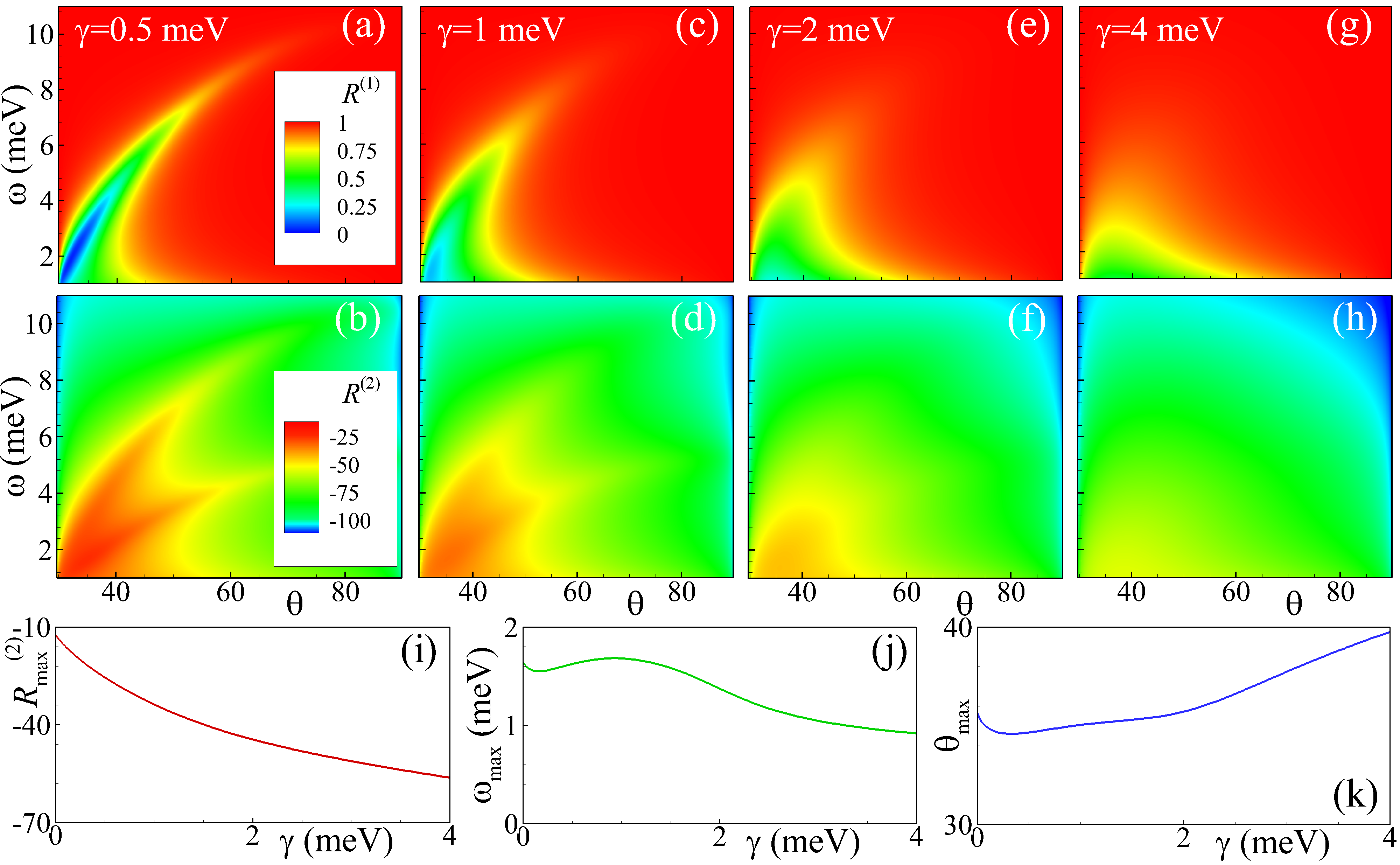}
		\caption{(a)--(h) Reflectance $R^{(1)}$ [top row, (a), (c), (e), and (g)] and SHG efficiency $R^{(2)}$ [middle row, (b), (d), (f), and (h)] vs angle of incidence $\theta$ and frequency $\omega$ of ATR structure for different inverse relaxation times $\gamma=0.5\,$meV (a), (b), $\gamma=1\,$meV (c), (d), $\gamma=2\,$meV (e), (f), $\gamma=4\,$meV (g), (h). (i)--(k) Dependence of the maximal SHG efficiency $R_{\rm max}^{(2)}$ (i), frequency $\omega_{\rm max}$ (j), and angle of incidence $\theta_{\rm max}$ (k) on inverse relaxation time $\gamma$. In all panels, other parameters are $d=10\,\mu$m, $\varepsilon_{3}=16$, $\varepsilon_{2}=1$, $\varepsilon_{1}=3.9$, $E_{F}=0.5\,$eV, $E_{i}=0.1\,$MV/cm.}
		\label{fig:R12-gamma}
	\end{figure*}

	The dependence of reflectance and SHG efficiency on the inverse scattering time is shown in Fig.~\ref{fig:R12-gamma}. Thus, with an increase of $\gamma$, the quality factor of the reflectance resonance gradually lowers [compare Figs.~\ref{fig:R12-gamma}(a), \ref{fig:R12-gamma}(c), \ref{fig:R12-gamma}(e), and \ref{fig:R12-gamma}(g)], while the SHG becomes less efficient [compare Figs.~ \ref{fig:R12-gamma}(b), \ref{fig:R12-gamma}(d), \ref{fig:R12-gamma}(f), and \ref{fig:R12-gamma}(h)]. Moreover, the maximal value of SHG efficiency drops dramatically from $\sim-10\,$dB to $\sim-60\,$dB when $\gamma$ is increased in the range $0\,$meV$<\gamma<4\,$meV [see Figs.~\ref{fig:R12-gamma}(i)].
	
	\bibliographystyle{apsrev4-2-titles}
	\bibliography{second_harmonics_bib}
	
\end{document}